\documentclass[
twocolumn,
aps,
prd,
superscriptaddress,
floatfix,
longbibliography
]{revtex4-1}
\usepackage[dvipdfmx]{graphicx}
\usepackage{dcolumn}
\usepackage{bm}
\usepackage{amsmath}
\usepackage{amssymb}
\usepackage{color} 
\usepackage{xcolor}
\usepackage{mathrsfs} 
\usepackage{algorithm}
\usepackage{algorithmicx}
\usepackage{algpseudocode}

\usepackage{hyperref}
\hypersetup{
  colorlinks=true,
  linkcolor=[rgb]{0.00,0.80,0.0},
  citecolor=[rgb]{0.0,0.0,1.0},
  urlcolor=[rgb]{1.0,0.0,0.4},
  setpagesize=false
}

\makeatletter
\newsavebox{\@brx}
\newcommand{\llangle}[1][]{\savebox{\@brx}{\(\m@th{#1\langle}\)}%
  \mathopen{\copy\@brx\kern-0.5\wd\@brx\usebox{\@brx}}}
\newcommand{\rrangle}[1][]{\savebox{\@brx}{\(\m@th{#1\rangle}\)}%
  \mathclose{\copy\@brx\kern-0.5\wd\@brx\usebox{\@brx}}}
\makeatother

\begin{document}

\title{Local multiplet formation around a single vacancy in graphene: an effective Anderson model analysis based on 
the block-Lanczos DMRG method}

\author{Tomonori Shirakawa} 
\affiliation{Computational Materials Science Research Team, RIKEN Center for Computational Science (R-CCS), Kobe, Hyogo 650-0047, Japan}
\affiliation{Quantum Computational Science Research Team, RIKEN Center for Quantum Computing (RQC), Wako, Saitama 351-0198, Japan}

\author{Seiji Yunoki}
\affiliation{Computational Materials Science Research Team, RIKEN Center for Computational Science (R-CCS), Kobe, Hyogo 650-0047, Japan}
\affiliation{Quantum Computational Science Research Team, RIKEN Center for Quantum Computing (RQC), Wako, Saitama 351-0198, Japan}
\affiliation{Computational Quantum Matter Research Team, RIKEN Center for Emergent Matter Science (CEMS), Wako, Saitama 351-0198, Japan}
\affiliation{Computational Condensed Matter Physics Laboratory, RIKEN Cluster for Pioneering Research (CPR), Saitama 351-0198, Japan}

\date{\today}

\begin{abstract}
To better understand the electronic structure of a single vacancy in graphene, 
we study the ground state property of an effective Anderson model, consisting of 
three dangling $sp^2$ orbitals of the surrounding carbon atoms around the vacancy  
and the $\pi$ orbitals of carbon atoms that form the honeycomb lattice with 
a single vacancy. 
This model possesses $\mathcal{C}_3$ point group symmetry around the vacancy 
and thereby the local multiplets can be characterized by their irreducible representations. 
Employing the block-Lanczos density-matrix renormalization group (DMRG) scheme proposed by the present authors 
[T. Shirakawa and S. Yunoki, Phys. Rev. B {\bf 90}, 195109 (2014)], we show that 
there are two phases in the relevant parameter space, i.e., a nonmagnetic phase 
in the weak coupling region and a free magnetic moment phase 
in the realistic parameter region. The systematic analysis finds that, 
in the free magnetic moment phase, local multiplets of the doubly degenerate 
irreducible representation with spin 1 become dominant in the ground state, 
and approximately half of this local spin 1 is screened by electrons in the surrounding 
$\pi$ orbitals, indicating the emergence of the residual spin-1/2 free magnetic moment. 
The symmetry of this local multiplets is compatible with the occurrence of 
the in-plane Jahn-Teller distortion to lift the degeneracy, found in the previous 
{\it ab-initio} calculations based on the density functional theory. 
Furthermore, we find that the emergence of the free magnetic moment is robust against 
carrier doping, which is in sharp contrast to the case of graphene with an adatom, thus 
explaining the qualitative difference observed experimentally in these two classes of systems. 
We also find the enhancement of the spin correlation function between $\pi$ electrons 
around the vacancy and those 
in the conduction band away from the vacancy in the undoped case, as compared to that in the doped case, 
while the spin correlation function between the $\sigma$ electrons 
in the $sp^2$ dangling orbitals around the vacancy and the $\pi$ electrons in the conduction band 
remains large in both undoped and doped cases. 
This implies that there is an additional contribution for the free magnetic moment 
from the $\pi$ electrons, which is fragile against the carrier doping, 
besides the free magnetic moment due to 
the $\sigma$ electrons in the $sp^2$ dangling orbitals, 
which is robust against the carrier doping.
Our calculations thus support qualitatively the previous experiment that suggests the
emergence of free magnetic moment with two distinct origins. 
\end{abstract}

\maketitle


\section{\label{sec:introduction}Introducion}

Graphene with vacancies has attracted a great deal of attention 
because it provides a route for additional functionality of graphene, i.e., the emergence of 
magnetism as predicted by {\it ab-initio} calculations based on the density-functional theory 
(DFT)~\cite{yazyev,dharma-wardana,dai,paz,padmanabhan,valencia},  
similar to the case of graphite~\cite{el-barbary,lehtinen}. 
Indeed, the magnetization measurements 
have observed paramagnetism in the presence of vacancies~\cite{ney,nair1,nair2}, 
while the pristine graphene is diamagnetic~\cite{sepioni}.
Moreover, the tunnel scanning microscope measurements have 
observed the spectra showing two spin-polarized peaks located in the vicinity of vacancy~\cite{yzhang}.

The introduction of a single vacancy in graphene induces dangling orbitals that are localized around the vacancy. 
These orbitals would contribute to reconstruct the electronic structure around the vacancy 
because the energy levels of these dangling orbitals are located near the Fermi level. 
Therefore, the correlation effect is expected to play an important role to 
determine the electronic structure even in the carbon based system.

The previous {\it ab-initio} DFT calculations have shown that 
there occurs the in-plane Jahn-Teller distortion around the vacancy~\cite{yazyev,dharma-wardana,dai,paz,padmanabhan},
in which two of three carbon atoms around the vacancy become closer in distance. 
Let us briefly consider the spatial symmetry of graphene in the presence of a single vacancy. 
If one of carbon atoms forming the honeycomb lattice structure is removed from graphene, 
the system has $\mathcal{C}_3$ point group symmetry around the vacancy. 
Under the $\mathcal{C}_3$ rotational symmetry, the eigenstates of the system are characterized with 
the non-degenerate symmetric irreducible representation $\mathcal{A}$ or the doubly degenerate 
irreducible representation $\mathcal{E}$.
The distortion found in the {\it ab-initio} DFT calculations is considered as the consequence of 
the coupling through the vibration of $\mathcal{E}$-type, called $e$ vibration mode. 
The electronic state coupled to this vibration mode is expected to be a state in $\mathcal{E}$ symmetry~\cite{casartelli}.  
This can indeed be compatible, for example, with a noninteracting state composed of the three $sp^2$-dangling orbitals 
around the vacancy occupied by three electrons.
However, if we consider a noninteracting system composed of three $\pi$ orbitals around the vacancy
in addition to the three $sp^2$-dangling orbitals,
there are six electrons and the ground state can be in $\mathcal{A}$ symmetry.
Furthermore, it is not trivial how the correlation effect along with the surrounding other orbitals 
can be incorporate in these pictures.

Another issue is the size of the free magnetic moment induced by the presence of vacancy in graphene.
It has been reported that the magnetization curve of graphene in the presence of vacancy generated
by the proton irradiation is well described by the Brillouin function with $1.0 \mu_B$~\cite{nair1,nair2} ($\mu_B$ being the Bohr magneton), 
while the other experiment for graphene nanoflakes irradiated with nitrogen ions suggests 
$2.0 \mu_B$ paramagnetism induced per single vacancy~\cite{ney}. 
Since graphene with vacancy is a complex system to analyze experimentally, 
the theoretical investigation based on the numerical approaches is highly anticipated to be useful.  
However, the estimated values by the {\it ab-initio} DFT calculations are 
rather widely distributed from $1.0\mu_B$ to $2.0 \mu_B$~\cite{yazyev,dharma-wardana,dai,paz,padmanabhan}, 
depending on the use of different types of the exchange potential~\cite{valencia} 
as well as the treatment of the further corrections for the electron correlation~\cite{casartelli,pinheiro}. 
Note also that the size of the free magnetic moment observed experimentally could not be simply estimated theoretically 
by the static one-body approximation because it fails to describe multiplet electronic structures and Kondo-like physics.   
Therefore, it is valuable to address these issues based on the many-body model calculations.

Moreover, other experiments suggest that the emergence of the free magnetic moment 
around the vacancy is robust against carrier doping~\cite{nair2,yzhang}.  
This is in sharp contrast to the magnetism induced by hydrogen absorption,  
where the magnetic moment vanishes by carrier doping~\cite{nair2,gonzalez-herrerd}. 
It is also claimed in Ref.~\cite{nair2} that the vacancy magnetism in graphene has a dual origin and 
approximately half of the free magnetic moment abruptly diminishes with carrier doping. 
Therefore, the mechanism for the emergence of the free magnetic moment in graphene  
would be different between these two systems with vacancies and adatoms.

To better understand the origin of magnetism, here we examine the ground state property of 
an effective Anderson model for a single vacancy in graphene. Although there 
have been several models proposed for graphene in the presence of vacancy~\cite{kanao,cazalilla,mitchell1}, 
the essential difference of our model is to contain all three $sp^2$ dangling orbitals around the vacancy,
in addition to the $\pi$ orbitals of carbon atoms that form the honeycomb lattice with a single vacancy. 
We also impose the $\mathcal{C}_3$ rotational symmetry around the vacancy 
to discuss a possible mechanism for the Jahn-Teller distortion.

First, we analyze the local multiplet structures around a vacancy based on a three-site two-orbital Hubbard model, 
composed of the three dangling $sp^2$ orbitals and three $\pi$ orbitals closest to the vacancy, 
which will be incorporated into an effective Anderson model 
as the impurity sites. 
We show that there are two phases in a reasonable parameter region: 
one is a weak coupling phase in which the multiplet structures are obtained 
by the perturbation theory to the open shell electron configuration realized in the noninteracting limit, 
and the other is a strong coupling phase in which each of carbon sites forms spin triplet 
due to the Hund's coupling.
The effective model in the strong coupling region is thus well described by the spin-1 antiferromagnetic 
Heisenberg model whose ground state is often referred to as a valence bond state~\cite{haldane,aklt}.
In both phases, the low-lying multiplet structures are identified by the irreducible representations 
of the $\mathcal{C}_3$ rotational point group and the total spin.

We then examine how the local multiplet ground state is affected via the hybridization 
with the surrounding $\pi$ electrons in the conduction band. 
For this purpose, we construct an effective Anderson model composed of the three-site two-orbital 
Hubbard model, which describes the local electronic states around the vacancy 
and thus serves as the impurity sites, 
and the surrounding $\pi$ orbitals of carbon atoms forming the honeycomb lattice with the single vacancy. 
The Anderson model is solved  
by employing the density matrix renormalization group (DMRG) method~\cite{white1,white2} 
combining with the block-Lanczos technique~\cite{shirakawa1,allerdt1,allerdt2,allerdt3},
which gives the numerically accurate solution within controlled errors.  
With this numerical analysis, we can fully take into account the local multiplet structures as well as the coupling to 
the surrounding $\pi$ electrons, 
which thus provides the valuable information complementary to the one-body type approximation 
used in the {\it ab-initio} DFT calculations.

Main results are summarized as follows.  
We find that (1) in the reasonable parameter region for graphene, the local multiplets with $\mathcal{E}$ 
irreducible representation and local spin $S=1$ are dominant in the ground state, 
due to the coupling with the surrounding $\pi$ electrons in the conduction band.  
The symmetry of the local $\mathcal{E}$ multiplets is thus compatible 
to the occurance of the in-plane Jahn-Teller distortion by the coupling through the 
$\mathcal{E}$-type vibration mode~\cite{casartelli}. 
We also find that (2) the local magnetic moment is not completely screened 
but partially screened by the surrounding $\pi$ electrons in 
the conduction band, suggesting the existence of the residual free magnetic moment. 
The estimated local free magnetic moment around the vacancy is as large as $\sim 0.55$ for the undoped case 
and $\sim 0.45$ for the doped case, where the g~factor $g = 2$ is assumed, 
implying the $S=1/2$ paramagnetism formed locally around the vacancy. 
We also confirm that (3) this free magnetic moment is robust against carrier doping,
which is compatible to the experimental results~\cite{nair2,yzhang}. 
Finally, analyzing the spin correlation function between the electrons in the local impurity sites
and the surrounding $\pi$ electrons in the conduction band, 
we find that (4) the spin correlation functions between the electrons in the dangling $sp^2$ orbitals 
and the surrounding $\pi$ electrons are insensitive to the carrier doping, 
while those between the electrons in the local $\pi$ orbitals around the vacancy and the surrounding 
$\pi$ electrons behave differently in the undoped and doped cases. 
The latter is attributed to the nature of the hybridization function that shows a pseudo-gap structure 
in the undoped case, thus leading to the unscreening of the local magnetic moment,
but does not in the doped case.

The rest of this paper is organized as follows. 
We first introduce an effective Anderson model for a single vacancy in graphene in Sec.~\ref{sec:model}. 
We then summarize, as the building blocks of the Anderson model,  
the low-lying multiplet structures of the local impurity part of the Anderson model, 
corresponding to an effective three-site two-orbital 
Hubbard model, and the hybridization function that describes the energy-resolved coupling 
between the local multiplets and the surrounding $\pi$ orbitals forming the conduction band of the Anderson model. 
The numerical method used in this study is briefly explained in Sec.~\ref{sec:method} and  
the numerical results are shown in Sec.~\ref{sec:results}. 
The paper is concluded in Sec.~\ref{sec:discussion} by discussing the relevance of our results to the experiments 
and also making the comparison with the previous theoretical studies. 
The explicit form of the local multiplet states in the strong coupling limit 
is provided in Appendix~\ref{app:localmultiplet}.
The spin correlation functions for similar but much simpler models are calculated 
as the reference for the comparison in Appendix~\ref{app:refspin}.

\section{Model}\label{sec:model}

\subsection{Anderson model for a single vacancy in graphene}

Graphene is composed of carbon atoms forming a two-dimensional honeycomb lattice structure 
with a single carbon atom locating at each site. 
The $sp^2$ orbitals of a carbon atom in graphene are three local orbitals 
formed by the linear combination of the 2$s$ atomic orbital and two 2$p$ atomic orbitals 
(2$p_x$ and 2$p_y$ orbitals) in the plane. 
Each of these $sp^2$ orbitals has a large lobe pointing to one of three different directions and 
the angle between any two of them is $120^{\circ}$, hence 
compatible to the honeycomb lattice structure of graphene. 
In graphene, two of these $sp^2$ orbitals in neighboring sites point to each other (one from each site) and  
hybridize strongly (i.e., $\sigma$ bonding) to form bonding and antibonding orbitals between the neighboring sites.  
Because of the large energy level difference between the bonding and antibonding orbitals, the bonding orbital 
is fully occupied by two electrons and the antibonding orbital is 
completely empty. Therefore, these orbitals formed by the $sp^2$ orbitals are usually inactive in the pristine graphene. 
The electronic properties of graphene are thereby determined 
by the remaining $\pi$ orbitals, i.e., $2p_z$ orbitals perpendicular to the graphene plane, 
and the resulting energy band exhibits the Dirac linear dispersion across the Fermi level~\cite{novoselov}. 
The $sp^2$ orbitals are referred to simply as $\sigma$ orbitals hereafter.

\begin{figure}[bhtp]
\begin{center}
\includegraphics[width=\hsize]{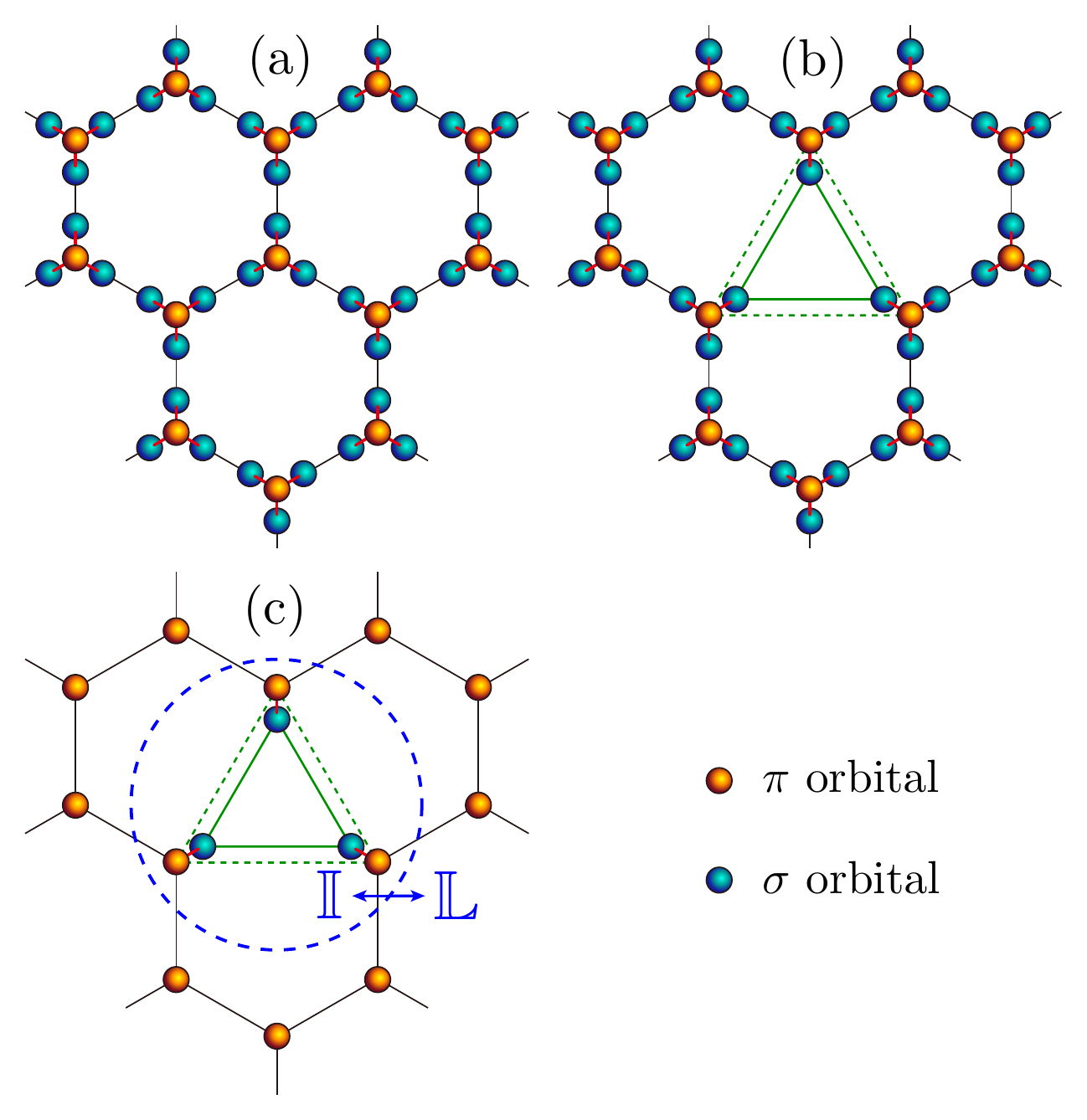}
\caption{
(a) Schematic depiction of the pristine graphene  
composed of $\pi$ orbitals (orange spheres) and $\sigma$ orbitals (blue spheres). 
Each carbon site has one $\pi$ orbital and three $\sigma$ orbitals. 
Here, $\sigma$ orbitals refer to the $sp^2$ orbitals and their directional nature of the shape of orbitals 
is indicated by locating them slightly away from the lattice point toward the neighboring sites.  
(b) Schematic depiction of graphene in the presence of a single vacancy. 
Three dangling $\sigma$ orbitals appear around the vacancy. 
(c) Schematic depiction of the effective Anderson model $\mathcal{H}_{\rm AM}$ studied here. 
The model consists of the three $\sigma$ orbitals around the vacancy and all $\pi$ orbitals 
of carbon atoms forming the honeycomb lattice structure with a single vacancy. 
Three sites around the vacancy are treated as an ``impurity site" in the Anderson model 
(inside a blue dashed circle denoted as $\mathbb{I}$) 
and the remaining sites are treated as the conduction sites 
(outside a blue dashed circle denoted as $\mathbb{L}$).  
Green solid (dashed) lines in (b) and (c) indicate the bonds inside region $\mathbb{I}$ 
with a finite  transfer integral $t_{\sigma}$ ($t_{\pi}$) between 
the $\sigma$ orbitals ($\pi$ orbitals). 
}
\label{fig:model}
\end{center}
\end{figure}

These features can be depicted schematically in Fig.~\ref{fig:model}(a) with orange (blue) spheres representing 
$\pi$ ($\sigma$) orbitals. 
Each carbon atom has six electrons. Two of them occupy the 1s core orbital and the remaining four electrons 
occupy the $\pi$ and $\sigma$ orbitals with one electron per orbital.  
All of the $\sigma$ orbitals have their neighboring pairs to form the covalent $\sigma$ bonding. 

Now, let us consider to introduce a single vacancy of the carbon atom in graphene. 
As shown in Fig.~\ref{fig:model}(b), in the presence of a single vacancy, 
there appear three $\sigma$ orbitals around the vacancy, which have no neighboring pairs. 
These unpaired $\sigma$ orbitals are called dangling orbitals and  
are tightly localized around the vacancy. 
On the other hand, the removal of one site 
generates the imbalance between sublattices in the honeycomb lattice~\cite{lieb}, 
which induces a zero energy mode in the $\pi$ orbital system~\cite{pereira,peres,ugeda}. 
The spectral weight (or equivalently the wave function) of the zero energy mode is mostly found 
around the vacancy, although it should be more extended in space than the dangling $\sigma$ orbitals.

For the modeling of this system, i.e., a single vacancy in graphene, we introduce the following simplifications. 
First, as shown in Fig.~\ref{fig:model}(c), we neglect all paired $\sigma$ orbitals
because they are inactive. 
Next, we assume that the conduction electrons in the surrounding $\pi$ orbitals away from the vacancy 
are described by a noninteracting tight-binding model, 
implicitly considering 
the renormalization effect of the correlated two-dimensional massless Dirac electrons~\cite{otsuka,seki}. 
Furthermore, we ignore possible modification of the hopping matrix elements around the vacancy. 
Within these assumptions, one can model our system as the following effective Anderson model: 
\begin{equation}
\mathcal{H}_{\rm AM} = \mathcal{H}_{\mathbb{I}} + \mathcal{H}_{\mathbb{L}} \label{eq:am},
\end{equation}
where $\mathcal{H}_{\mathbb{I}}$ and $\mathcal{H}_{\mathbb{L}}$ describe the local part around the vacancy 
and the surrounding part of the system away from the vacancy, respectively [see Fig.~\ref{fig:model}(c)].

The local part of the Hamiltonian, $\mathcal{H}_{\mathbb{I}}$, is composed of the three dangling $\sigma$ orbitals 
and the three $\pi$ orbitals on the three carbon sites next to the vacancy: 
\begin{equation}
\mathcal{H}_{\mathbb{I}} = \mathcal{H}_{t} + \mathcal{H}_U + \mathcal{H}_d, \label{eq:hloc}
\end{equation}
where 
\begin{align}
\mathcal{H}_{t} & = 
\varepsilon_{\sigma} 
\sum_{i \in \mathbb{I} } \sum_{s=\uparrow,\downarrow} c_{i\sigma s}^{\dagger} c_{i\sigma s} + 
\varepsilon_{\pi} \sum_{i \in \mathbb{I}} \sum_{s=\uparrow,\downarrow} c_{i\pi s}^{\dagger} c_{i\pi s} \nonumber \\
& - t_{\sigma} 
\sum_{\scriptsize \llangle i,j \rrangle_{\mathbb{I}}} \sum_{s=\uparrow,\downarrow} c_{i\sigma s}^{\dagger} c_{j\sigma s} 
- t_{\pi} \sum_{\scriptsize \llangle i,j \rrangle_{\mathbb{I}}}
\sum_{s=\uparrow,\downarrow} c_{i\pi s}^{\dagger} c_{j\pi s} \label{eq:hop},
\end{align}
\begin{align}
\mathcal{H}_U & = U \sum_{i \in \mathbb{I}} \sum_{\alpha=\sigma,\pi} n_{i\alpha\uparrow} n_{i\alpha\downarrow} 
+ U^{\prime} \sum_{i \in \mathbb{I}} n_{i\sigma} n_{i\pi} \nonumber \\
& - 2 J \sum_{i \in \mathbb{I}} \left( {\bm S}_{i\sigma} \cdot {\bm S}_{i\pi} + \frac{1}{4} n_{i\sigma} n_{i\pi} \right) \nonumber \\
& + J^{\prime} \sum_{i \in \mathbb{I}} 
\left( 
c_{i\sigma\uparrow}^{\dagger} c_{i\sigma\downarrow}^{\dagger} c_{i\pi\downarrow} c_{i\pi\uparrow} +
c_{i\pi\uparrow}^{\dagger} c_{i\pi\downarrow}^{\dagger} c_{i\sigma\downarrow} c_{i\sigma\uparrow}
\right), \label{eq:int}
\end{align}
and 
\begin{equation}
\mathcal{H}_d = - \frac{1}{2}(U+2U^{\prime}-J) \sum_{i \in \mathbb{I}} \sum_{\alpha=\sigma,\pi} n_{i\alpha}.  \label{eq:dc}
\end{equation}
Here, $\mathbb{I}$ represents the set of site labels $i$ for the three carbon sites next to the vacancy [see Fig.~\ref{fig:model}(c)], 
$\llangle i,j \rrangle_{\mathbb{I}}$ denotes 
a pair of sites $i$ and $j$ in $\mathbb{I}$, 
corresponding to the next nearest neighbors in the honeycomb lattice. 
$c_{i\alpha s}$ ($c_{i\alpha s}^{\dagger}$) is the annihilation (creation) operator 
of an electron at site $i$ with spin $s$ ($=\uparrow,\downarrow$) and orbital $\alpha$ (=$\sigma$, $\pi$), 
$n_{i \alpha s} = c_{i\alpha s}^{\dagger} c_{i\alpha s}$, and $n_{i \alpha} = n_{i \alpha \uparrow} + n_{i \alpha \downarrow}$. 
${\bm S}_{i\alpha}=(S_{i\alpha}^x, S_{i\alpha}^y, S_{i\alpha}^z)$ is 
the local spin operator for orbital $\alpha$ at site $i$ given by 
\begin{equation}
S_{i\alpha}^{\nu} = \frac{1}{2} \sum_{s,s^{\prime}=\uparrow,\downarrow} 
c_{i\alpha s}^{\dagger} [ \mbox{\boldmath{$\sigma$}}^{\nu} ]_{s s^{\prime}} c_{i\alpha s^{\prime}}, \quad 
(\nu=x,y,z)
\label{eq:spinoperator}
\end{equation}
where $\mbox{\boldmath{$\sigma$}}^x$, $\mbox{\boldmath{$\sigma$}}^y$, and $\mbox{\boldmath{$\sigma$}}^z$ 
represent the Pauli matrices. 
We also define the total spin operator at site $i$,  
\begin{equation}
{\bm S}_{i} = {\bm S}_{i\sigma} + {\bm S}_{i\pi} \label{eq:localspin}
\end{equation}
with ${\bm S}_{i}=(S_i^x, S_i^y, S_i^z)$, and the total spin operator over all sites in $\mathbb{I}$, 
\begin{equation}
{\bm S}_{\mathbb{I}} = \sum_{i \in \mathbb{I}} {\bm S}_i \label{eq:totalspin}
\end{equation}
with ${\bm S}_{\mathbb{I}}  = (S_{\mathbb{I}}^x, S_{\mathbb{I}}^y, S_{\mathbb{I}}^z)$. 

The first two terms in $\mathcal{H}_t$ represent the local energy levels of the $\sigma$ and $\pi$ orbitals. 
The last two terms in $\mathcal{H}_t$ are the hopping terms between different sites. 
Note that there is no hopping between the two different orbitals because the $\sigma$ orbital and the $\pi$ orbital are 
orthogonal to each other. 
For simplicity, we parametrize $t_{\sigma} = 2 t_{\pi}$ throughout this paper. 
$\mathcal{H}_U$ represents the local interaction terms, including 
the intra-orbital Coulomb interaction $U$, 
the inter-orbital Coulomb interaction $U^{\prime}$, 
the Hund's coupling $J$, and the pair hopping $J^{\prime}$. 
To reduce the number of parameters, here we assume $U=U^{\prime}+2J$, $J=J^{\prime}$, and $J=0.1U$, 
although the former is valid only for atomic orbitals~\cite{kanamori}. 
Notice also that $\mathcal{H}_d$ is added to correct the double counting of the interactions. 
While $\mathcal{H}_d$ does not make any difference 
when the isolated $\mathcal{H}_{\mathbb{I}}$ is considered, 
it becomes important to compensate the electron density 
when the surrounding $\pi$ orbitals described by $\mathcal{H}_{\mathbb{L}}$ is incooporated.

In the Anderson model $\mathcal{H}_{\rm AM}$ in Eq.~(\ref{eq:am}), 
the effects of the surrounding $\pi$ orbitals are encapsulated 
by the noninteracting Hamiltonian, 
\begin{align}
\mathcal{H}_{\mathbb{L}} = & - t \sum_{\langle i,j \rangle^{\prime}} \sum_{s=\uparrow,\downarrow} 
( c_{i\pi s}^{\dagger} c_{j \pi s} + c_{j\pi s}^{\dagger} c_{i \pi s} ) \nonumber \\
& - t_{\pi} \sum_{\mbox{\scriptsize{$\llangle$}} i,j \mbox{\scriptsize{$\rrangle$}}^{\prime}} \sum_{s=\uparrow,\downarrow} 
( c_{i\pi s}^{\dagger} c_{j \pi s} + c_{j\pi s}^{\dagger} c_{i \pi s} ), \label{eq:bath}
\end{align}
where $\langle i,j \rangle^{\prime}$ and $\llangle i,j \rrangle^{\prime}$ 
indicate pairs of nearest-neighbor and next-nearest-neighbor sites $i$ and $j$, respectively, on the honeycomb lattice 
except for pairs of sites inside $\mathbb{I}$, implying that pairs of sites across the boundary between $\mathbb{I}$ and 
$\mathbb{L}$ are also included.  
As indicated in Fig.~\ref{fig:model}(c), 
in the follows, we simply use $\mathbb{L}$ to denote the set of all sites 
in the honeycomb lattice but excluding sites $i \in \mathbb{I}$, i.e., 
all sites in the surrounding $\pi$ orbital system.

\subsection{\label{sec:multiplet}Local multiplets}

Setting aside the surrounding $\pi$ orbital system, 
we shall first examine the low-lying multiplet structure 
of the local part of the Hamiltonian $\mathcal{H}_{\mathbb{I}}$ given in Eq.~(\ref{eq:hloc}), 
which is essentially the three-site two-orbital Hubbard model. 
The local part of the Hamiltonian $\mathcal{H}_{\mathbb{I}}$ can be considered as the building block, 
i.e., ``impurity site", of the Anderson model $\mathcal{H}_{\rm AM}$ and the understanding of the low-lying multiplet structure 
is essential to understand the electronic structure of the total system. 
Since each orbital has nominally one electron in the effective model of graphene considered here, 
we impose that the total number of electrons in $\mathcal{H}_{\mathbb{I}}$ is six.

The model described by $\mathcal{H}_{\mathbb{I}}$ in Eq.~(\ref{eq:hloc}) possesses the $\mathcal{C}_3$ point group symmetry 
as well as the spin rotational symmetry. Therefore, any eigenstate of $\mathcal{H}_{\mathbb{I}}$ can be 
simultaneously characterized with the irreducible representation of the point group and the total spin. 
As already described in Sec.~\ref{sec:introduction}, 
the irreducible representations of the $\mathcal{C}_3$ point group are the symmetric one-dimensional 
representation $\mathcal{A}$ and the doubly degenerate representation $\mathcal{E}$. 
In this paper, we denote sets of states belonging to these irreducible representations $\mathcal{A}$ and 
$\mathcal{E}$ as $\mathbb{A}$ and $\mathbb{E}$, respectively. 
We also represent the quantum numbers of ${\bm S}_{\mathbb{I}}$ and 
$S_{\mathbb{I}}^z$ as $S_{\mathbb{I}}$ and $M_{\mathbb{I}}$, respectively. 
Namely, the eigenvalue of ${\bm S}_{\mathbb{I}} \cdot {\bm S}_{\mathbb{I}}$ is $S_{\mathbb{I}}(S_{\mathbb{I}}+1)$. 
Using these notations, we can refer to the energy eigenstates with ($\mathbb{C}$, $S_{\mathbb{I}}$) 
($\mathbb{C}=\mathbb{A}$, $\mathbb{E}$) as the multiplets
characterized by the irreducible representation $\mathcal{C}$ 
($\mathcal{C}=\mathcal{A}$, $\mathcal{E}$) and the total spin $S_{\mathbb{I}}$.

\begin{figure}[htbp]
\begin{center}
\includegraphics[width=\hsize]{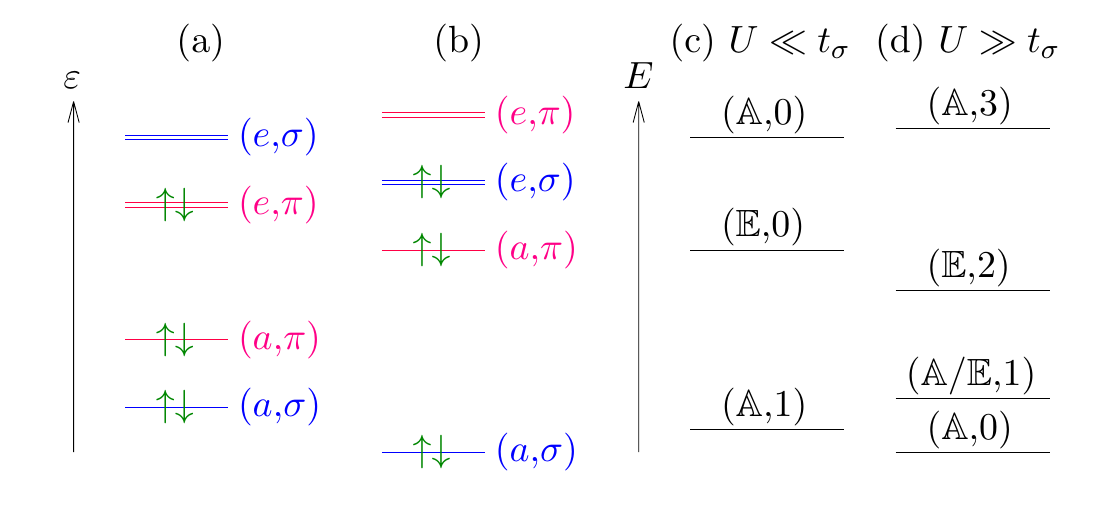}
\caption{ 
(a, b) Single-particle energy level diagrams and the ground-state electron configurations
of $\mathcal{H}_{\mathbb{I}}$ in the noninteracting limit 
for (a) $-t_{\sigma} < \varepsilon_{\pi}-\varepsilon_{\sigma} < t_{\sigma}/2$ 
and (b) $t_{\sigma}/2 < \varepsilon_{\pi}-\varepsilon_{\sigma} < 2t_{\sigma}$ with $t_{\sigma}=2t_{\pi}>0$. 
$(a,\alpha)$ [($e,\alpha$)] for orbital $\alpha\, (= \sigma,\pi)$ characterizes each single-particle energy level  
with the irreducible representation $\mathcal{A}$ ($\mathcal{E}$) of the $\mathcal{C}_3$ point group. 
Green arrows represent up and down electrons. 
(c, d) Low-lying multiplet energy diagrams in (c) the weak coupling limit ($U \ll t_{\sigma}$)
and (d) the strong coupling limit ($U \gg t_{\sigma}$). 
$(\mathbb{C},S_{\mathbb{I}})$ ($\mathbb{C}=\mathbb{A},\mathbb{E}$, $S_{\mathbb{I}}=0,1,2,3$) 
characterizes the associated many-body energy eigenstates, i.e., multiplets, with 
the irreducible representation $\mathcal{X}$ ($\mathcal{X}=\mathcal{A}$, $\mathcal{E}$) and the total spin $S_{\mathbb{I}}$. 
Here the total number of electrons is assumed to be 6. 
}
\label{fig:hloc}
\end{center}
\end{figure}

In the noninteracting case, the single-particle energy level $\varepsilon_{\alpha}$ for each orbital $\alpha\,(=\sigma,\pi)$ 
is split by the electron hopping $t_\alpha$ into a nondegenerate energy level with $\mathbb{A}$ and 
doubly degenerate energy levels with $\mathbb{E}$. 
We denote these single-particle energy levels as $a \in \mathbb{A}$ and $e \in \mathbb{E}$ with their energy levels 
$\varepsilon_{\alpha}^{(a)}$ and $\varepsilon_{\alpha}^{(e)}$, respectively. 
These energy levels $\varepsilon_{\alpha}^{(a)}$ and $\varepsilon_{\alpha}^{(e)}$ are easily evaluated as 
$\varepsilon_{\sigma}^{(a)}=-2t_{\sigma}+ \varepsilon_{\sigma}$ and $\varepsilon_{\sigma}^{(e)}=t_{\sigma} + \varepsilon_{\sigma}$ 
for the $\sigma$ orbitals, and $\varepsilon_{\pi}^{(a)}=-2t_{\pi} + \varepsilon_{\pi}$ and 
$\varepsilon_{\pi}^{(e)}=t_{\pi} + \varepsilon_{\pi}$ for the $\pi$ orbitals. 
For $t_{\sigma} = 2 t_{\pi}>0$ and $-t_{\sigma} < \varepsilon_{\pi} - \varepsilon_{\sigma} < t_{\sigma}/2$, 
the ground-state electron configuration with six electrons
is open shell with two electrons occupying the $e$ level of $\pi$ orbitals, 
as shown in Fig.~\ref{fig:hloc}(a). 
Notice that in this case the number of electrons in the $\sigma$ ($\pi$) orbitals, 
$N_{\sigma}$ ($N_{\pi}$), is $N_{\sigma} = 2$ ($N_{\pi} = 4$). 
 The first order perturbation theory with respect to the Coulomb interactions lifts the six-fold degenerate  
electron configurations of the open shell $e$ level of $\pi$ orbitals. 
Consequently, the ground state becomes ($\mathbb{A}, 1$), as shown in Fig.~\ref{fig:hloc}(c). 
When $t_{\sigma}/2 < \varepsilon_{\pi} - \varepsilon_{\sigma} < 2 t_{\sigma}$ 
is realized together with $t_{\sigma} = 2 t_{\pi}>0$, 
the $e$ level of $\sigma$ orbitals is now lower than the $e$ level of $\pi$ orbitals, thus 
indicating that $N_{\sigma}=4$ and $N_{\pi}=2$, as shown in Fig.~\ref{fig:hloc}(b). 
However, the resulting ground state in this case is again ($\mathbb{A}, 1$) in the weak coupling limit, 
as shown in Fig.~\ref{fig:hloc}(c).

In the strong coupling limit, the Coulomb interactions force the electron distribution to be uniform, 
yielding $N_{\sigma} = N_{\pi} = 3$. Moreover, the Hund's coupling $J$ makes the local state a spin triplet 
($S_i=1$) at each site. Since the kinetic exchange mechanism induces the antiferromagnetic interactions 
between neighboring sites, the low-energy effective model for $\mathcal{H}_{\mathbb{I}}$ 
in the strong coupling limit is represented by the three-site spin-1 antiferromagnetic Heisenberg model. 
The low-lying energy levels for this effective model are summarized in Fig.~\ref{fig:hloc}(d). 
The ground state is ($\mathbb{A}, 0$), which refers to a kind of the valence-bond-type state~\cite{aklt} 
in the spin-1 antiferromagnetic chain~\cite{haldane}. More detailed analysis of the strong coupling limit is 
provided in Appendix~\ref{app:localmultiplet}.

\begin{figure}[htbp]
\begin{center}
\includegraphics[width=\hsize]{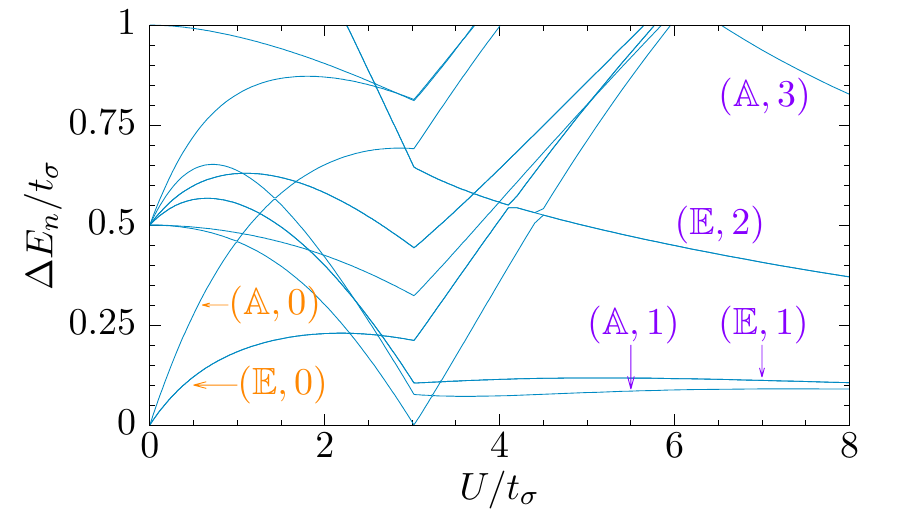}
\caption{
Low-lying energy $\Delta E_n=E_n-E_0$ of $\mathcal{H}_{\mathbb{I}}$ as a function of $U$. 
Here, $E_n$ is the $n$th energy eigenvalues of $\mathcal{H}_{\mathbb{I}}$ with $n=0$ corresponding to the ground state and 
$E_0\le E_1\le E_2\le \cdots$. 
$(\mathbb{C},S_{\mathbb{I}})$ ($\mathbb{C}=\mathbb{A},\mathbb{E}$, $S_{\mathbb{I}}=0,1,2,3$) is indicated for some 
of the energy eigenvalues $E_n$ to characterize the corresponding energy eigenstates. 
The parameters used here are $t_{\sigma}=2t_{\pi}>0$, $\varepsilon_{\sigma}=\varepsilon_{\pi}$, and $J/U=0.1$. 
The ground state ($n=0$) is $(\mathbb{A},1)$ for $U/t_\sigma <3$ and $(\mathbb{A},0)$ for $U/t_\sigma>3$.
The total number of electrons is assumed to be 6.
}
\label{fig:hloc2}
\end{center}
\end{figure}

In order to confirm the multiplet energy diagrams shown in Figs.~\ref{fig:hloc}(c) and \ref{fig:hloc}(d), 
we show in Fig.~\ref{fig:hloc2} the low-lying energy $\Delta E_n$ defined as  
\begin{equation}
\Delta E_n = E_n - E_0,   
\end{equation}
where $E_n$ ($n=0,1,2,\cdots$) is the $n$th energy eigenvalues of $\mathcal{H}_{\mathbb{I}}$ with six electrons 
and $n=0$ corresponds to the ground state energy, i.e., $E_0\le E_1\le E_2\le \cdots$, 
calculated numerically by fully diagonalizing the Hamiltonian $\mathcal{H}_{\mathbb{I}}$.   
Characterizing the energy eigenstates, 
we find that the low-lying multiplet structure is precisely reproduced by 
those shown in Fig.~\ref{fig:hloc}(c) in the weak coupling limit ($U \approx 0$) and in Fig.~\ref{fig:hloc}(d) in 
the strong coupling limit ($U \gg t_{\sigma}$), 
except for the small energy splitting between the multiplets with $(\mathbb{A},1)$ and $(\mathbb{E},1)$, 
which is expected to be degenerate in the limit of $U/t_\sigma\to\infty$. 
Notice also in Fig.~\ref{fig:hloc2} that the transition from the weak coupling limit to the strong coupling limit, 
occurring at $U=3 t_{\sigma}$ for the specific parameter used here, 
is essentially a level crossing. 
Indeed, these two states with $(\mathbb{A},1)$ and $(\mathbb{A},0)$ are distinguishable with different eigenvalues 
of the total spin $S_{\mathbb{I}}$.

\begin{figure}[htbp]
\begin{center}
\includegraphics[width=\hsize]{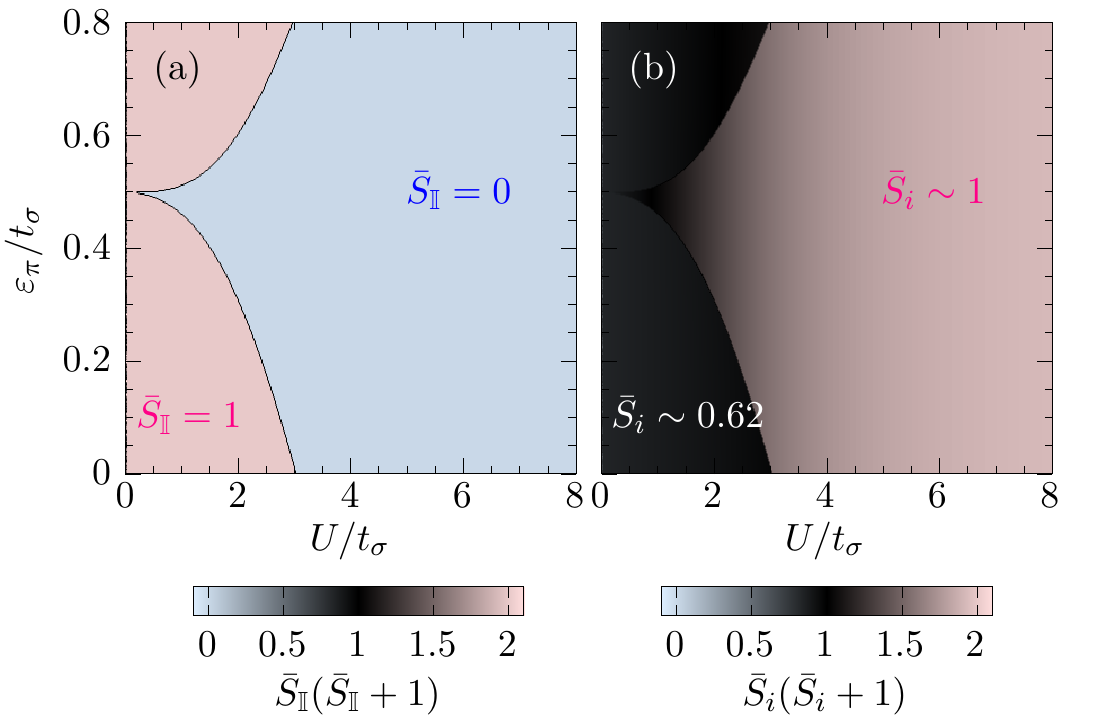}
\caption{
Intensity plots of (a) $\langle {\bm S}_{\mathbb{I}} \cdot {\bm S}_{\mathbb{I}} \rangle\equiv \bar{S}_{\mathbb{I}}(\bar{S}_{\mathbb{I}}+1)$ 
and (b) $\langle {\bm S}_{i} \cdot {\bm S}_{i} \rangle\equiv \bar{S}_i (\bar{S}_i + 1)$ calculated for the ground state of 
$\mathcal{H}_{\mathbb{I}}$ with $t_{\sigma}=2t_{\pi}>0$, $\varepsilon_{\sigma}=\varepsilon_{\pi}$, and $J/U=0.1$. 
The values of $\bar{S}_{\mathbb{I}}$ and $\bar{S}_i$ are also indicated in (a) and (b), respectively. 
The total number of electrons is assumed to be 6.
}
\label{fig:hloc3}
\end{center}
\end{figure}

Figure \ref{fig:hloc3}(a) shows the intensity plot of 
$\langle {\bm S}_{\mathbb{I}} \cdot {\bm S}_{\mathbb{I}} \rangle\equiv \bar{S}_{\mathbb{I}}(\bar{S}_{\mathbb{I}}+1)$ 
obtained by numerically diagonalizing $\mathcal{H}_{\mathbb{I}}$ with six electrons. 
Here, $\langle \mathcal{O} \rangle$ indicates the expectation value of an operator $\mathcal{O}$ 
for the ground state. 
As expected, the total spin of the ground state for the small $U/t_\sigma$ region is $\bar{S}_{\mathbb{I}}=1$ 
and that for the large $U/t_\sigma$ region is $\bar{S}_{\mathbb{I}}=0$. 
We also find in Fig.~\ref{fig:hloc3}(b) that, for the large $U/t_\sigma$ region, 
the total spin at each site, $\bar{S}_{i}$, 
evaluated from $\langle {\bm S}_{i} \cdot {\bm S}_{i} \rangle\equiv \bar{S}_i (\bar{S}_i + 1)$, 
approaches to $\bar{S}_i=1$ with increasing $U/t_\sigma$, 
suggesting the formation of a local spin 1 at each site, where the low-lying states 
can be described by the spin-1 antiferromagnetic Heisenberg model.
It is also clear in Fig.~\ref{fig:hloc3}(a) that the phase boundary between the weak and strong coupling phases 
corresponds to a level crossing 
because the total spin of the ground state is different.

There are two remarks in order. 
First, any multiplet state in both weak and strong coupling phases 
cannot be described by a single Slater determinant, 
indicating the importance of the correlation effect. 
Especially, the Hund's coupling play an essential role on realizing 
the valence-bond-type state in the strong coupling phase. 
Second, $t_{\pi}$ is the next nearest neighbor hopping for the $\pi$ orbitals and is estimated 
as large as $t_{\pi} \sim t/12$, where $t$ is the nearest neighbor hopping for the $\pi$ orbitals in graphene~\cite{kretinin}. 
On the other hand, the intra-orbital Coulomb interaction for the $\pi$ orbitals is estimated 
as large as $\sim 3.5t$ by the constrained random phase approximation~\cite{wehling}. 
Therefore, it is reasonable to expect that the ground state of $\mathcal{H}_{\mathbb{I}}$
with the parameter set for graphene is in the strong coupling phase. 
However, as we will discuss below, the effect of the surrounding $\pi$ orbitals is dominant  
to determine the low-lying electronic structure of the Anderson model $\mathcal{H}_{\rm AM}$.

\subsection{\label{sec:hyb}Hybridization function}

It is known that an Anderson model in general is characterized by a local Hamiltonian for the impurity sites, 
which corresponds to $\mathcal{H}_{\mathbb{I}}$ in our case, 
and a so-called hybridization function $\Gamma_{ij}(\omega)$~\cite{bulla}, mathematically defined 
as the Schur complement of the inverse of the hopping matrix (i.e., Green's function)
for the complementary space of the impurity sites in $\mathbb{I}$, which corresponds to 
$\mathbb{L}$ in our case~\cite{shirakawa1}. 
Physically, $\Gamma_{ij}(\omega)$ is a quantity characterizing the energy resolved coupling 
of the impurity sites to the surrounding bath sites, i.e., the surrounding $\pi$ orbital system in $\mathbb{L}$. 
For the Anderson model described by the Hamiltonian $\mathcal{H}_{\rm AM}$ given in Eq.~(\ref{eq:am}), 
$\Gamma_{ij}(\omega)$ is a three-by-three matrix labeled 
by the site indices of the $\pi$ orbitals in $\mathbb{I}$, 
since the $\sigma$ orbitals in $\mathbb{I}$ do not hybridize with the surrounding $\pi$ orbital system in $\mathbb{L}$. 
Because of the $\mathcal{C}_3$ point group symmetry of the system, $\Gamma_{ij}(\omega)$ 
can be diagonalized with the diagonal elements
$\tilde{\Gamma}_{a}(\omega)$ and $\tilde{\Gamma}_{e}(\omega)$
corresponding to the couplings with the $(a,\pi)$~orbital and the $(e,\pi)$~orbitals ($e=e_1,e_2$) in $\mathbb{I}$, respectively. 
We call the $(a,\pi)$-orbital system $a$-mode and the $(e,\pi)$-orbital systems $e$-modes.

\begin{figure}[htbp]
\begin{center}
\includegraphics[width=\hsize]{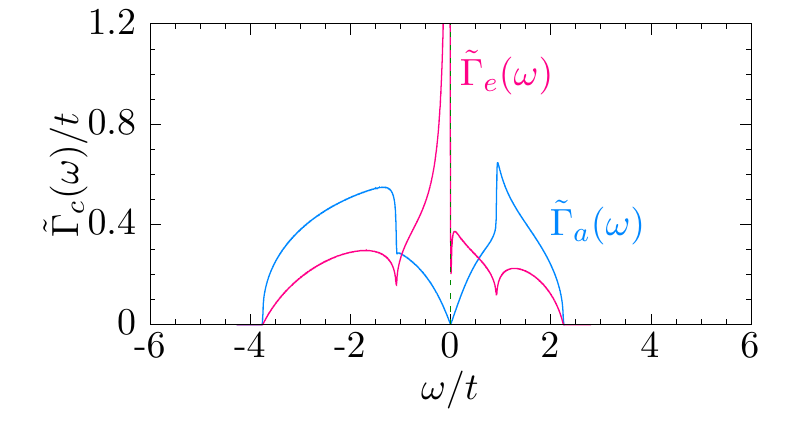}
\caption{Hybridization functions $\tilde{\Gamma}_{c}(\omega)$ ($c=a,e$)
for $\mathcal{H}_{\rm AM}$ with $t_{\pi}=t/12$. 
Green dashed line indicates the Fermi level at $\omega=0$ for bulk graphene, i.e., undoped graphene. 
Note that hybridization functions $\tilde{\Gamma}_{e}(\omega)$ for $e=e_1$ and $e_2$ are exactly the same 
and here only one of them is plotted.
}
\label{fig:dhyb}
\end{center}
\end{figure}

Figure \ref{fig:dhyb} shows $\tilde{\Gamma}_{a}(\omega)$ and $\tilde{\Gamma}_{e}(\omega)$ 
for the Anderson model $\mathcal{H}_{\rm AM}$ given in Eq.~(\ref{eq:am}) with $t_{\pi}=t/12$~\cite{kretinin}.
Around the Fermi level for bulk graphene, i.e., undoped graphene, a diverging behavior appears in $\tilde{\Gamma}_e(\omega)$, 
which is associated with the zero energy modes of the $\pi$ orbital system~\cite{pereira,peres}, as described below. 
This diverging behavior strongly affects the distribution of the local multiplets 
coupled to the $e$-modes, as shown later. 
We should note here that because of the particle-hole asymmetry due to the presence of the next-nearest hopping $t_{\pi}$,  
$\tilde{\Gamma}_e(\omega)$ does not exactly diverge and the maximum location in energy is slightly shifted below  
the Fermi level~\cite{pereira}.
In this case, there are two phases expected for the ground state, according to the previous studies~~\cite{mitchell1}.
One is the symmetric strong coupling phase where the Kondo screening state arises.
The other is the asymmetric strong coupling phase where the spin degree of freedom is vanished
because of the large potential difference between the impurity and conduction sites that induces the closed-shell configuration 
at the impurity site.
The divergent hybridization function is preferable to the symmetric strong coupling 
even when the model is particle-hole asymmetric.

On the other hand, we find in Fig.~\ref{fig:dhyb} that $\tilde{\Gamma}_a(\omega)$ shows 
a pseudogap structure at the Fermi level. 
An Anderson model with a single magnetic impurity coupled to the bath sites 
through a hybridization function with a pseudogap structure at the Fermi level is 
called a pseudogap Anderson model~\cite{fritz}. 
The pseudogap Anderson model is indeed considered as an effective model 
for a hydrogen adatom on graphene~\cite{shirakawa1,shirakawa2}.
Previous studies have revealed that the Kondo screening is absent 
in the pseudogap Anderson model~\cite{gonzalez-buxton,vojta}. 
Consequently, there appears the free magnetic moment even at zero temperature. In other words, 
the hybridization function with the pseudogap structure is irrelevant 
to the low-lying electronic properties of the impurity site. 
The carrier doping into the pseudogap Anderson model 
yields either the Kondo screening of the magnetic moment at the impurity site or 
the vanishment of the magnetic moment by adding or removing electrons at the impurity site~\cite{vojta}, 
suggesting that the free magnetic moment in the adatom graphene 
would be sensitive against the carrier doping.

Finally, we note that the diverging part in $\tilde{\Gamma}_{e}(\omega)$ 
is sometimes referred to as the zero energy modes~\cite{pereira,peres,ugeda}. 
The number of the zero energy modes~\cite{pereira,peres,ugeda} is related to 
the imbalance of the number of sites in the two sublattices of a bipartite system~\cite{lieb}. 
If we consider the system $\mathbb{L}$, 
the imbalance of the number of sites in the two sublattices is $2$ ($=3-1$), 
which coincides with the degeneracy of $\mathcal{E}$. 
From this rule, we can readily understand that $\tilde{\Gamma}_{e}(\omega)$ should show the diverging behavior
because of the presence of the two zero energy modes,
while $\tilde{\Gamma}_{a}(\omega)$ should not exhibit any diverging behavior that originates from 
the zero energy modes caused by the sublattice imbalance.

\section{\label{sec:method}Method}

\begin{figure}
\begin{center}
\includegraphics[width=\hsize]{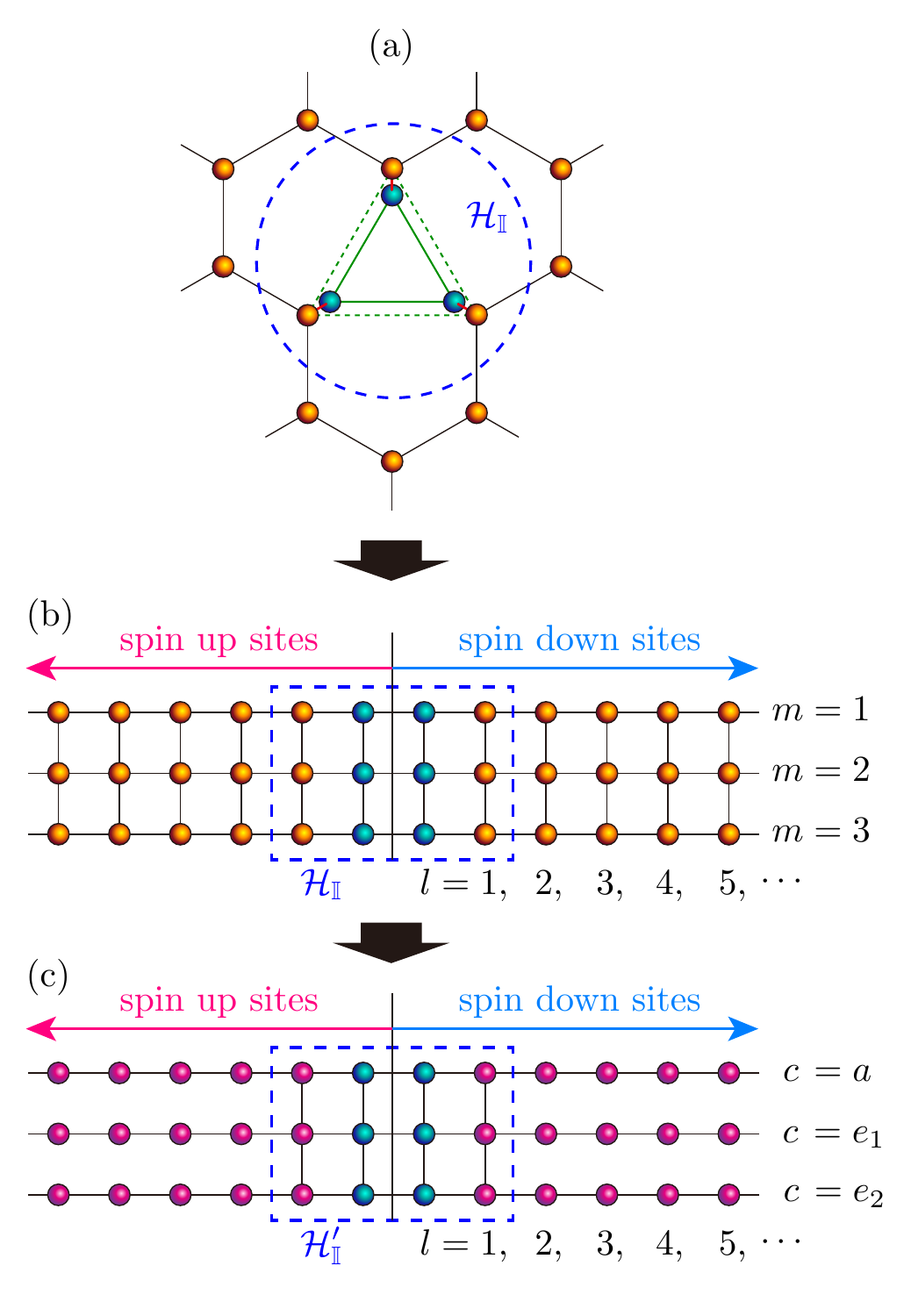}
\caption{
Schematic depictions of (a) the Anderson model $\mathcal{H}_{\rm AM}$ given in Eq.~(\ref{eq:am}) for a single vacancy in 
graphene, (b) the effective Q1D three-leg Anderson model $\mathcal{H}_{\rm Q1D}$ obtained by 
the block Lanczos transformation with the $\pi$ orbitals in $\mathbb{I}$ as the initial block Lanczos bases, 
and (c) the symmetrized Q1D three-leg Anderson model 
$\mathcal{H}_{\rm Q1D}^{\prime}$ obtained by the unitary transformation 
in Eqs.~(\ref{eq:q1d:symbase:a})--(\ref{eq:q1d:symbase:e2}).
$m$ ($=1,2,3$) and $c$ ($=a,e_1,e_2$) stand for labels of legs in the Q1D ladder representations, and 
$l$ stands for a label of rungs corresponding to the block Lanczos basis generated by the block Lanczos iteration. 
Blue spheres in (a)--(c) denote the $\sigma$ orbitals in $\mathbb{I}$ represented by $(c_{i\sigma s}, c_{i\sigma s}^\dag)$. 
Orange and red spheres in (a), (b), and (c) denote the $\pi$ orbitals in both $\mathbb{I}$ and $\mathbb{L}$ 
represented by $(c_{i\pi s}, c_{i\pi s}^\dag)$, $(\tilde{c}_{lm s}, \tilde{c}_{lm s}^\dag)$, and 
$(\tilde{f}_{l\pi s}, \tilde{f}_{l\pi s}^\dag)$, respectively. 
Note that $l=1$ corresponds to the initial bases for the block Lanczos iteration and thus $(\tilde{c}_{lm s}, \tilde{c}_{lm s}^\dag)$ 
for $l=1$ in (b) are exactly the same as $(c_{i\pi s}, c_{i\pi s}^\dag)$ for $i\in\mathbb{I}$. 
The region indicated by the blue dashed rectangle corresponds to the local Hamiltonian 
$\mathcal{H}_{\mathbb{I}}$, which remains unchanged under the transformation in (b). 
However, the form of $\mathcal{H}_{\mathbb{I}}$ is changed in (c). 
The left (right) part of the sites in (b) and (c) is assigned to represent orbitals with up spin 
(down spin). 
}
\label{fig:method}
\end{center}
\end{figure}

We use the density-matrix renormalization group (DMRG) method~\cite{white1,white2} to study the ground state properties 
of the Anderson model described by the Hamiltonian $\mathcal{H}_{\rm AM}$ in Eq.~(\ref{eq:am}). 
For this purpose, we 
employ the block Lanczos tridiagonalization technique to map the Anderson model $\mathcal{H}_{\rm AM}$ 
onto an effective quasi-one-dimensional (Q1D) model~\cite{shirakawa1,allerdt1,allerdt2,allerdt3}, 
which consists of the three $\sigma$ orbitals and the three $\pi$ orbitals in $\mathbb{I}$, describing the local part of the Hamiltonian 
$\mathcal{H}_{\mathbb{I}}$, 
and $3(L-1)$ $\pi$ orbitals in $\mathbb{L}$, generated by the block Lanczos procedure 
starting 
from the three $\pi$ orbitals in $\mathbb{I}$, to represent the surrounding $\pi$ orbital system that couples to the $\pi$ orbitals 
in $\mathbb{I}$ [see Figs.~\ref{fig:method}(a) and \ref{fig:method}(b)]. 
Here, $L$ is the maximum number of the block Lanczos iterations. 
With the local part of the Hamiltonian $\mathcal{H}_{\mathbb{I}}$ in Eqs.~(\ref{eq:hloc})--(\ref{eq:dc}),
the Hamiltonian $\mathcal{H}_{\rm Q1D}$ of the resulting effective Q1D model is thus given by 
\begin{align}
\mathcal{H}_{\rm Q1D} & = \mathcal{H}_{\mathbb{I}} + \sum_{s=\uparrow,\downarrow} \mathcal{H}^{s}_c \label{eq:q1dmodel0}, 
\end{align}
where 
\begin{align}
\mathcal{H}^{s}_c = & \sum_{l=1}^L \sum_{m=1}^3 \sum_{m^{\prime}=1}^3 
[ \mbox{\boldmath{$\varepsilon$}}_l ]_{m,m^{\prime}} \tilde{c}_{lm s}^{\dagger} \tilde{c}_{lm^{\prime} s} \nonumber \\
& + \sum_{l=1}^{L-1} \sum_{m=1}^3 \sum_{m^{\prime}=1}^3 [ \mbox{\boldmath{$\tau$}}_l ]_{m,m^{\prime}} 
\tilde{c}_{l m s}^{\dagger} \tilde{c}_{l+1 m^{\prime} s} + {\rm h.c.} \label{eq:q1dladder}
\end{align}
and $\tilde{c}_{lm s}$ ($m=1,2,3$, $l=1,2,\cdots,L$) is an electron annihilation operator  
defined by the linear combination of $c_{i\pi s}$ in $\mathcal{H}_{\rm AM}$ through 
\begin{equation}
\tilde{c}_{l m s} = \sum_{i} U^{l}_{i,m} c_{i\pi s} \label{eq:q1dtransform}.
\end{equation}
Note that $\tilde{c}_{l m s}$ ($m=1,2,3$) for $l=1$ is identical to $c_{i \pi s}$ for $i \in \mathbb{I}$ and therefore 
the form of $\mathcal{H}_{\mathbb{I}}$, including the interaction part $\mathcal{H}_U$, remains unchanged under this transformation.

Let us now introduce the matrix $[ {\bm U} ]_{i,3l+m} \equiv [ {\bm U}_l ]_{i,m} \equiv U^l_{i,m}$ 
with $m=1,2,3$, $l=1,2,\cdots,L$, and $i \in \mathbb{I}+\mathbb{L}$ in Eq.~(\ref{eq:q1dtransform}). 
To obtain the form of the Hamiltonian $\mathcal{H}_c^s$
in Eq.~(\ref{eq:q1dladder}), we should notice that the block Lanczos transformation is applied to the surrounding part of 
the Hamiltonian $\mathcal{H}_{\mathbb{L}}$ defined in Eq.~(\ref{eq:bath}).
The matrix ${\bm U}$ represents the block Lanczos basis that transforms 
the Hamiltonian matrix of $\mathcal{H}_{\mathbb{L}}$ for the $\pi$ orbitals with spin $s$, i.e.,  
\begin{equation}
[ {\bm H}_{\mathbb{L}}^s ]_{ij} = \left\{ 
\begin{array}{ll}
-t & i,j \in \langle i,j \rangle \\
-t_{\pi} & i,j \in \llangle i,j \rrangle \\
0 & \text{otherwise} \\
\end{array} 
\right.
\end{equation}
into a tri-block-diagonal matrix form, i.e.,  
\begin{equation}
{\bm U}^{\dagger} {\bm H}_{\mathbb{L}}^s {\bm U} = \left( 
\begin{array}{cccccc}
\mbox{\boldmath{$\varepsilon$}}_1 & \mbox{\boldmath{$\tau$}}_1 & {\bm 0} & {\bm 0} & \cdots & {\bm 0} \\
\mbox{\boldmath{$\tau$}}_1^{\dagger} & \mbox{\boldmath{$\varepsilon$}}_2 & \mbox{\boldmath{$\tau$}}_2 & {\bm 0} & \cdots & {\bm 0} \\
{\bm 0} & \mbox{\boldmath{$\tau$}}_2^{\dagger} & \mbox{\boldmath{$\varepsilon$}}_3 & \mbox{\boldmath{$\tau$}}_3 & {} & \vdots \\
{\bm 0} & {\bm 0} & \mbox{\boldmath{$\tau$}}_3^{\dagger} & \ddots & \ddots & {} \\
\vdots & \vdots & {} & \ddots & \mbox{\boldmath{$\varepsilon$}}_{L-1} & \mbox{\boldmath{$\tau$}}_{L-1} \\
{\bm 0} & {\bm 0} & \cdots & {} & \mbox{\boldmath{$\tau$}}_{L-1}^{\dagger} & \mbox{\boldmath{$\varepsilon$}}_{L} \\
\end{array}
\right), \label{eq:tridiagmat}
\end{equation}
where $\mbox{\boldmath{$\varepsilon$}}_l$ and $\mbox{\boldmath{$\tau$}}_l$ are the $3 \times 3$ matrices appearing 
in Eq.~(\ref{eq:q1dladder}).
We can obtain these matrices recursively through the block Lanczos iteration 
\begin{equation}
{\bm U}_{l+1} \mbox{\boldmath{$\tau$}}_{l+1}^{\dagger} = {\bm H}_{\mathbb{L}}^s {\bm U}_l - {\bm U}_l \mbox{\boldmath{$\varepsilon$}}_l - {\bm U}_{l-1} \mbox{\boldmath{$\tau$}}_{l},
\label{eq:blocklanczos}
\end{equation}
where $\mbox{\boldmath{$\varepsilon$}}_1=0$, 
${\bm U}_0=0$, ${\bm U}_1=(\bm{e}_{i_1},\bm{e}_{i_2},\bm{e}_{i_3})$ with $\bm{e}_{i}$ being a column unit vector, i.e., 
$(\bm{e}_i)_k=\delta_{i,k}$ and $\{ i_1, i_2, i_3 \} \equiv \mathbb{I}$,
and the left hand side is obtained by the QR decomposition of the right hand side. 
Note that the particular form of $\bm{U}_1$ is due to the initial block Lanczos bases chosen intentionally 
to be the three $\pi$ orbitals in $\mathbb{I}$.
In addition, as shown in Fig.~\ref{fig:method}(b), we can divide the system into two blocks using the spin degrees of freedom 
to reduce the local Hilbert space~\cite{shirakawa1}. 
More details of the derivation and the technical information for Eqs.~(\ref{eq:q1dmodel0})--(\ref{eq:blocklanczos}) 
can be found in Ref.~\cite{shirakawa1}.

Because of the $\mathcal{C}_3$ rotational symmetry, we can find that the matrix elements of $\mbox{\boldmath{$\varepsilon$}}_l$ 
and $\mbox{\boldmath{$\tau$}}_l$ satisfy
\begin{align}
& [ \mbox{\boldmath{$\varepsilon$}}_l ]_{11} = [ \mbox{\boldmath{$\varepsilon$}}_l ]_{22} = [ \mbox{\boldmath{$\varepsilon$}}_l ]_{33}, \label{eq:q1d:emat:diag} \\
& [ \mbox{\boldmath{$\varepsilon$}}_l ]_{12} = [ \mbox{\boldmath{$\varepsilon$}}_l ]_{23} = [ \mbox{\boldmath{$\varepsilon$}}_l ]_{31}
= [ \mbox{\boldmath{$\varepsilon$}}_l ]_{21} = [ \mbox{\boldmath{$\varepsilon$}}_l ]_{32} = [ \mbox{\boldmath{$\varepsilon$}}_l ]_{13}, \label{eq:q1d:emat:offdiag} \\
& [ \mbox{\boldmath{$\tau$}}_l ]_{11} = [ \mbox{\boldmath{$\tau$}}_l ]_{22} = [ \mbox{\boldmath{$\tau$}}_l ]_{33}, \label{eq:q1d:tmat:diag} 
\end{align}
and
\begin{align}
& [ \mbox{\boldmath{$\tau$}}_l ]_{12} = [ \mbox{\boldmath{$\tau$}}_l ]_{23} = [ \mbox{\boldmath{$\tau$}}_l ]_{31}
= [ \mbox{\boldmath{$\tau$}}_l ]_{21} = [ \mbox{\boldmath{$\tau$}}_l ]_{32} = [ \mbox{\boldmath{$\tau$}}_l ]_{13}. \label{eq:q1d:tmat:offdiag}
\end{align}
Therefore, the unitary transformation of the electron operators 
\begin{align}
\tilde{f}_{l a s} & = ( \tilde{c}_{l 1 s} + \tilde{c}_{l 2 s} + \tilde{c}_{l 3 s} )/\sqrt{3}, \label{eq:q1d:symbase:a} \\
\tilde{f}_{l e_1 s} & = ( 2 \tilde{c}_{l 1 s} - \tilde{c}_{l 2 s} - \tilde{c}_{l 3 s} )/\sqrt{6}, \label{eq:q1d:symbase:e1} 
\end{align}
and
\begin{align}
\tilde{f}_{l e_2 s} & = ( \tilde{c}_{l 2 s} - \tilde{c}_{l 3 s} )/\sqrt{2} \label{eq:q1d:symbase:e2} 
\end{align}
for $l=1,2,\cdots,L$ transforms $\mathcal{H}_{\rm Q1D}$ into a decoupled form (apart from the local part in $\mathbb{I}$) 
of the Q1D model 
\begin{align}
\mathcal{H}_{\rm Q1D}^{\prime} & = \mathcal{H}_{\mathbb{I}}^\prime + \sum_{s=\uparrow,\downarrow} \mathcal{H}^{s}_{f}, \label{eq:q1dmodel} 
\end{align}
where
\begin{align}
\mathcal{H}^{s}_{f} = & \sum_{l=1}^L \sum_{c=a,e_1,e_2} 
\tilde{\varepsilon}_{lc} \tilde{f}_{l c s}^{\dagger} \tilde{f}_{l c s} \nonumber \\
& + \sum_{l=1}^{L-1} \sum_{c=a,e_1,e_2} \tilde{\tau}_{lc} 
\tilde{f}_{l c s}^{\dagger} \tilde{f}_{l+1 c s} + {\rm h.c.}, \label{eq:q1dsymladder}
\end{align}
as shown schematically in Fig.~\ref{fig:method}(c). 
Note that 
because of the unitary transformation, the form of the local part of the Hamiltonian $\mathcal{H}_{\mathbb{I}}$, including the interaction 
part, is changed to $ \mathcal{H}_{\mathbb{I}}^\prime $. 
We should also note that $c=a$ ($c=e_1,e_2$) indeed corresponds to the orbitals 
with the $\mathcal{A}$ ($\mathcal{E}$) irreducible representation of the $\mathcal{C}_3$ point group.

The systems sizes studied here are up to $L=47$, i.e., $6(L+1) = 288$ sites in the Q1D model including the spin degrees of 
freedom, which corresponds to approximately $7000$ sites in the honeycomb lattice arround the vacancy~\cite{shirakawa1}. 
We keep the number $\chi$ of the density matrix eigenstates up to $\chi = 120(L+1)$ to obtain the reasonable convergence.
For example, we have checked that the results do not change much even if we increase the number of the density matrix eigenstates 
kept up to $\chi=144 (L+1)$, and a typical error of the ground state energy is estimated as large as $10^{-5}t$.

Finally, we must pay special attention to the treatment of the total number of electrons in this Q1D representation. 
In this study, the number of electrons, $N_e$, is determined so as to reproduce 
the chemical potential $\mu$ for the bulk system. 
For a given chemical potential $\mu$, 
the ground potential $\Omega (N, \mu) = E_0(N) - \mu N$ 
becomes minimum at $N_e$ if $N_e$ satisfies 
\begin{equation}
\frac{E_0(N_e) - E_0(N_e-2)}{2} < \mu < \frac{E_0(N_e+2)-E_0(N_e)}{2}, \label{eq:chem}
\end{equation}
where $E_0(N)$ is the ground state energy of the Q1D model with $N$ electrons. 
The condition (\ref{eq:chem}) may also be written as
\begin{equation}
N_e = \underset{N}{\mathrm{min}} \Omega (N,\mu) 
\label{eq:chem:cond}
\end{equation}
for a given $\mu$. 
To obtain a target chemical potential $\mu$ for a given set of parameters 
($\varepsilon_{\sigma}$, $\varepsilon_{\pi}$, $t_{\sigma}$, $t_{\pi}$, $U$, $J$, $L$), 
we search $N_e$ that satisfies the condition in Eq.~(\ref{eq:chem:cond}) 
by calculating $E_0(N)$ for several values of $N$.

\section{\label{sec:results}Results}

In this section, we provide the numerical results obtained by the block Lanczos DMRG method for 
the ground state of the Anderson model $\mathcal{H}_{\rm AM}$ in Eq.~(\ref{eq:am}). 
The results for undoped and doped cases are shown in Sec.~\ref{sec:undoped} and Sec.~\ref{sec:doped}, 
respectively. The spin correlation functions between the impurity and conduction sites are also examined 
in Sec.~\ref{sec:spin}. 
Here, we set the hopping parameters $t=6t_{\sigma}=12t_{\pi}$~\cite{kretinin},
although the realistic value for $t_{\sigma}$ would be larger 
because the dangling $\sigma$ orbitals should be expanded around the vacancy. 
We also set $\varepsilon_{\sigma} = \varepsilon_{\pi} = 0$.

\subsection{\label{sec:undoped}Undoped case}

As described at the end of the previous section, 
we first have to determine the number of electrons, $N_e$, in the Q1D model $\mathcal{H}_{\rm Q1D}$ 
so as to reproduce the chemical potential $\mu = 2 t_{\pi}$ of the undoped bulk system. 
To this end, we calculate the ground state energy $E_0(N_e)$ for a given 
$L$ with varying the number $N_e$ of electrons, and find that the condition 
$N_e : 3(L+1) = 7 : 6$ satisfies Eq.~(\ref{eq:chem:cond}) with $\mu = 2 t_{\pi}$ 
for the parameter region studied here. Note that $3(L+1)$ is the total number of sites 
in the Q1D model $\mathcal{H}_{\rm Q1D}$, excluding the spin degrees of freedom.

\begin{figure}
\begin{center}
\includegraphics[width=\hsize]{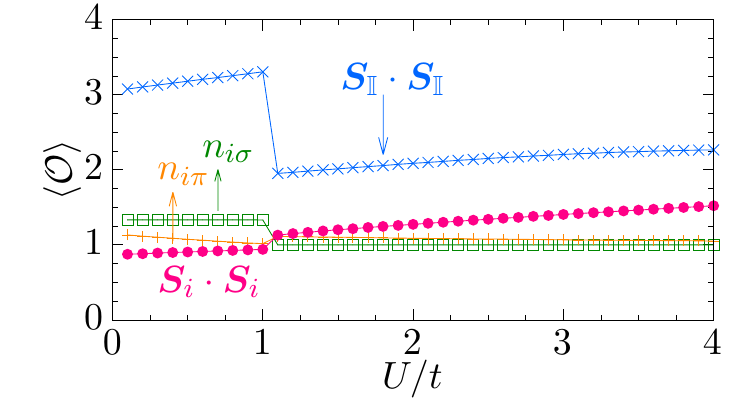}
\caption{ 
Expectation values $\langle \mathcal{O} \rangle$ of 
operators $\mathcal{O} = n_{i\sigma}$, $n_{i\pi}$, ${\bm S}_i \cdot {\bm S}_i$, and 
${\bm S}_{\mathbb{I}} \cdot {\bm S}_{\mathbb{I}}$, where $i\in\mathbb{I}$,
calculated for the ground state of the Anderson model $\mathcal{H}_{\rm AM}$. 
The parameters used here are $L=35$, $J=0.1U$, $t = 6 t_{\sigma} = 12 t_{\pi}$, and $\mu = 2 t_{\pi}$,
corresponding to the undoped case. 
}
\label{fig:dmrgn1166}
\end{center}
\end{figure}

Figure~\ref{fig:dmrgn1166} shows the $U$ dependence of the expectation values of local operators, 
including the local density $n_{i\alpha}$ of orbital $\alpha$ ($\alpha=\sigma,\pi$) at site $i \in \mathbb{I}$, 
the local spin squared ${\bm S}_i \cdot {\bm S}_i$ at site $i \in \mathbb{I}$, 
and the total spin squared ${\bm S}_{\mathbb{I}} \cdot {\bm S}_{\mathbb{I}}$ over sites 
$i \in \mathbb{I}$, for the ground state of the Anderson model $\mathcal{H}_{\rm AM}$. 
We find that there are two distinguishable phases: 
$\langle n_{i\sigma} \rangle = 4/3$ in a weak coupling region ($U \leq t$) 
and $\langle n_{i\sigma} \rangle = 1$ in a strong coupling region ($U > t$). 
We refer to these two phases as the 
weak coupling phase and the strong coupling phase, respectively.

The abrupt change of the expectation values at $U \approx t$ found in Fig.~\ref{fig:dmrgn1166} suggests 
that the transition from the weak to the strong coupling phase is of the first order due to a level crossing. 
This is because of the symmetry sector of the number of $\sigma$ electrons,  $N_{\sigma} = \sum_{i\in \mathbb{I}} n_{i\sigma}$. 
In our model, the symmetry sectors with even $N_{\sigma}$ and odd $N_{\sigma}$ are orthogonal to each other. 
Notice that the pair hopping term in $\mathcal{H}_{\mathbb{I}}$ 
changes $N_{\sigma}$ by $\pm2$ and thus does not mix the even and odd sectors. 
This also implies that the strong coupling phase cannot be accessible perturbatively from the noninteracting limit.

In the weak coupling phase, we find 
that $\langle {\bm S}_{\mathbb{I}} \cdot {\bm S}_{\mathbb{I}} \rangle\sim 3$, 
corresponding to $\bar{S}_{\mathbb{I}} \sim 1.3$,  
which is different from $S_{\mathbb{I}} = 1$ for the ground-state local multiplet $(\mathbb{A},1)$ in the weak coupling phase 
found in Fig.~\ref{fig:hloc}(c) and Fig.~\ref{fig:hloc3}(a). 
The deviation is attributed to the fact that the total number of 
electrons, $N_{\mathbb{I}}=\sum_{i\in\mathbb{I}}\sum_{\alpha=\sigma,\pi}n_{i\alpha}$, is larger than six. 
Indeed, we find here that 
$\langle N_{\mathbb{I}}\rangle = 3 (\langle n_{i\sigma} \rangle +  \langle n_{i\pi} \rangle) \sim 7.4$ 
for $U = 0.5 t$.

On the other hand, in the strong coupling phase, we find 
that $\langle {\bm S}_{\mathbb{I}} \cdot {\bm S}_{\mathbb{I}} \rangle \sim 2$, 
corresponding to $\bar{S}_{\mathbb{I}} \sim 1$, which is again different from 
$\bar{S}_{\mathbb{I}}=0$ for the ground-state local multiplet $(\mathbb{A},0)$  
in the strong coupling limit found in Fig.~\ref{fig:hloc}(d) and Fig.~\ref{fig:hloc3}(a). 
One should notice here that the total number $\langle N_{\mathbb{I}}\rangle$ of electrons is now as large as 6.
However, the expectation value of the local spin squared $\langle {\bm S}_i \cdot {\bm S}_i \rangle$ 
increases with $U$ and becomes as larger as 1.5 (i.e., $\bar{S}_i\sim0.82$) at $U = 4t$, 
suggesting the tendency to form the local site spin 1 with further increasing $U$, 
as expected in the strong coupling limit shown in Fig.~\ref{fig:hloc3}(b). 
Therefore, these results reveal that the dominant local multiplet structure found here in the ground state 
of $\mathcal{H}_{\rm AM}$ is different from the ground-state local multiplet $(\mathbb{A},0)$ 
shown in Fig.~\ref{fig:hloc}(d) and Fig.~\ref{fig:hloc3}(a) 
due to the effect of the surrounding $\pi$ electrons in the conduction band, which deserves more analysis.

\begin{figure}[htbp]
\begin{center}
\includegraphics[width=0.75\hsize]{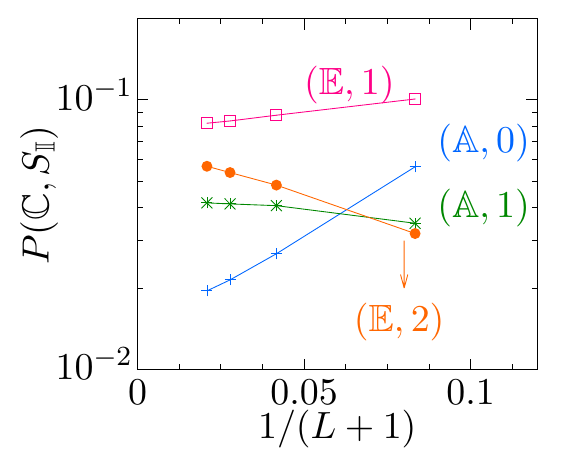}
\caption{ 
$L$ dependence of the projected weight $P(\mathbb{C}, S_{\mathbb{I}})$ of the 
ground state of the Anderson model $\mathcal{H}_{\rm AM}$ 
onto the local multiplet states $(\mathbb{C}, S_{\mathbb{I}})$ of $\mathcal{H}_{\mathbb{I}}$ in the strong coupling limit. 
Here, the ground state of the Anderson model $\mathcal{H}_{\rm AM}$ is obtained by mapping it to 
the Q1D model $\mathcal{H}_{\rm Q1D}$ with $L$ at $U=3t$ in the strong coupling phase. 
The other parameters used are $J=0.1U$, $t = 6 t_{\sigma} = 12 t_{\pi}$, and $\mu = 2 t_{\pi}$,
corresponding to the undoped case. 
}
\label{fig:dmrgn1166_2}
\end{center}
\end{figure}

Figure~\ref{fig:dmrgn1166_2} shows the projected weight $P(\mathbb{C}, S_{\mathbb{I}})$ of the ground state 
of the Anderson model $\mathcal{H}_{\rm AM}$ in the strong coupling phase onto the local multiplet states
$( \mathbb{C}, S_{\mathbb{I}} )$ with $\mathbb{C}=\mathbb{A},\mathbb{E}$ and $S_{\mathbb{I}}=0,1,2$ 
of $\mathcal{H}_{\mathbb{I}}$ in the strong coupling limit [see Fig.~\ref{fig:hloc}(d)].
Note that the projected weight for the local multiplet state $(\mathbb{A},3)$ is not shown in 
Fig.~\ref{fig:dmrgn1166_2} (and also in Fig.~\ref{fig:dmrgn1000_2}) because we find that it is quite small (less than $10^{-4}$).
The explicit form of the local multiplet states and the definition of the projected weight $P( \mathbb{C}, S_{\mathbb{I}} )$
are given in Eqs.~(\ref{eq:multipleta0s0})--(\ref{eq:multipleta0s3}) and Eq.~(\ref{eq:projectedweight}), respectively, 
in Appendix~\ref{app:localmultiplet}. 
It is clearly found in Fig.~\ref{fig:dmrgn1166_2} that the projected weight for $(\mathbb{E},1)$ is dominant 
over that for $(\mathbb{A},0)$ at large $L$, 
suggesting that the coupling to the surrounding $\pi$ electrons in the conduction band 
through the hybridization function 
$\tilde{\Gamma}_e(\omega )$ enhances the local multiplet component with $(\mathbb{E},1)$.

\begin{figure}[htbp]
\begin{center}
\includegraphics[width=0.75\hsize]{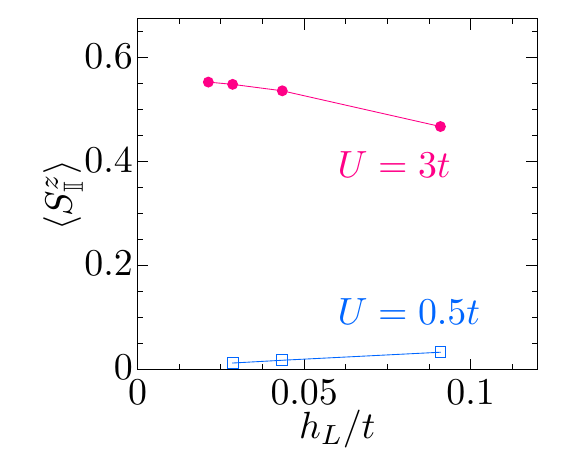}
\caption{ 
Finite size scaling of $\langle S_{\mathbb{I}}^z \rangle$ by applying an external magnetic field $h_L=t/L$ 
for the weak coupling phase at $U=0.5t$ and the strong coupling phase at $U=3t$. 
The other parameters are the same as in Fig.~\ref{fig:dmrgn1166_2}. 
}
\label{fig:dmrgn1166_3}
\end{center}
\end{figure}

To examine whether or not the local spin $\bar{S}_{\mathbb{I}}$ around the vacancy 
found in Fig.~\ref{fig:dmrgn1166} 
is screened by the surrounding $\pi$ electrons, we now calculate the expectation value of $S^z_{\mathbb{I}}$ 
for the ground state when an external magnetic field $h_L$ is applied on $\mathbb{I}$ 
in the Anderson model $\mathcal{H}_{\rm AM}$: 
\begin{equation}
\mathcal{H}_{\rm AM} \to \mathcal{H}_{\rm AM} - h_L S^z_{\mathbb{I}}.
\label{eq:h_hz}
\end{equation}
Here, the existence or absence of the local free magnetic moment in the thermodynamic limit can be analyzed by scaling 
the magnetic field $h_L = t/L$ with increasing $L$ when the ground state is calculated by mapping 
$\mathcal{H}_{\rm AM}$ into 
the Q1D model $\mathcal{H}_{\rm Q1D}$ with $L$.  
Figure~\ref{fig:dmrgn1166_3} shows the $L$ dependence of $\langle S_{\mathbb{I}}^z \rangle$. 
In the weak coupling phase ($U=0.5t$), we find that $\langle S_{\mathbb{I}}^z \rangle$ approaches to zero 
with decreasing $h_L$ (i.e., increasing the system size $L$), indicating that the local spin 
$\bar{S}_{\mathbb{I}} \sim 1.3$ around the vacancy is completely screened by the surrounding $\pi$ electrons. 
In contrast, in the strong coupling phase ($U=3t$), Fig.~\ref{fig:dmrgn1166_3} clearly shows that 
$\langle S_{\mathbb{I}}^z \rangle$ in the limit of $L \to \infty$ approaches to a finite value, as large as $0.55$,
indicating that the local spin $\bar{S}_{\mathbb{I}} \sim 1$ around the vacancy 
is partially screened by the surrounding $\pi$ electrons.

\subsection{\label{sec:doped}Doped case}

\begin{figure}[htbp]
\begin{center}
\includegraphics[width=\hsize]{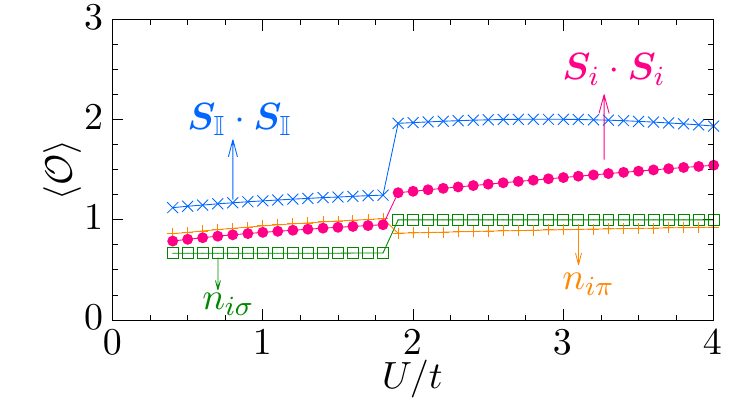}
\caption{
Same as Fig.~\ref{fig:dmrgn1166} except for $L=39$ and $N_e = 3(L+1)$, corresponding to 
the doped case. 
}
\label{fig:dmrgn1000}
\end{center}
\end{figure}

Let us now examine the ground state properties for the doped case. 
For this purpose, we set that the number $N_e$ of electrons is
$N_e = 3 (L+1)$ in the Q1D model $\mathcal{H}_{\rm Q1D}$ with $L$. 
This corresponds to a case of hole doping. 
The results of the local quantities are summarized in Fig.~\ref{fig:dmrgn1000}. 
As in the undoped case, we find that there are two distinguishable phases,  
the weak coupling phase and the strong coupling phase, separated at $U\approx1.8t$.  

\begin{figure}[htbp]
\begin{center}
\includegraphics[width=0.75\hsize]{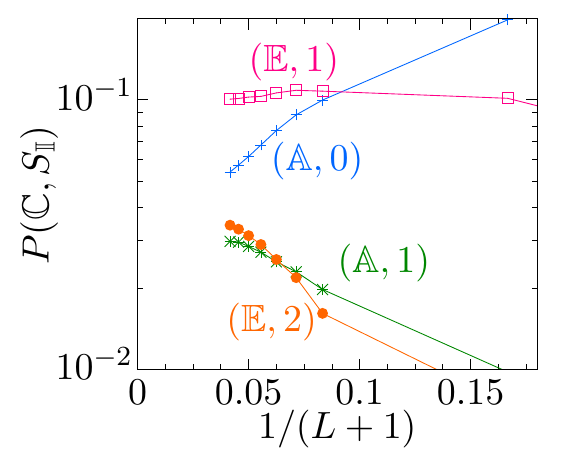}
\caption{
Same as Fig.~\ref{fig:dmrgn1166_2} except for $N_e = 3(L+1)$, corresponding to 
the doped case. 
}
\label{fig:dmrgn1000_2}
\end{center}
\end{figure}

\begin{figure}[htbp]
\begin{center}
\includegraphics[width=0.75\hsize]{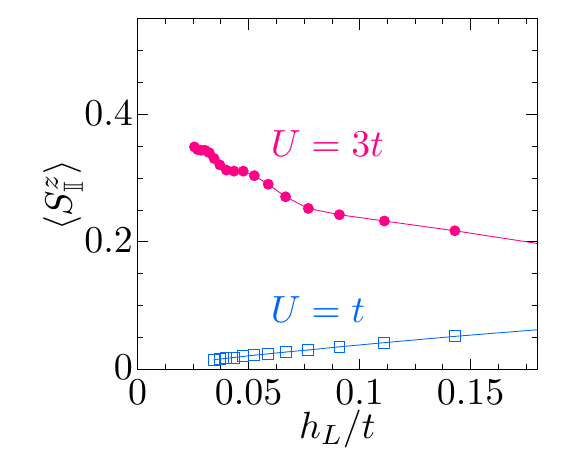}
\caption{
Same as Fig.~\ref{fig:dmrgn1166_3} except for $U=t$ in the weak coupling phase
and $N_e = 3(L+1)$, corresponding to the doped case. 
}
\label{fig:dmrgn1000_3}
\end{center}
\end{figure}

Figures~\ref{fig:dmrgn1000_2} and \ref{fig:dmrgn1000_3} show the $L$ dependence of the projected weight 
$P( \mathbb{C}, S_{\mathbb{I}})$ and $\langle S_{\mathbb{I}}^z \rangle$ induced by the external magnetic field $h_L=t/L$,
respectively, calculated at $U=3t$ in the strong coupling phase.  
These results correspond to Figs.~\ref{fig:dmrgn1166_2} and \ref{fig:dmrgn1166_3} for the undoped case.  
Comparing these results, we find that the overall features in the strong coupling phase are 
qualitatively the same as those for the undoped case.  
Namely, the emergence of the local free magnetic moment is robust under the carrier doping and the local spin 
$\bar{S}_{\mathbb{I}} \sim 1$ around the vacancy 
is only partially screened by the surrounding $\pi$ electrons. 
Indeed, the difference between the undoped and doped cases is rather qualitative. 
For example, the unscreened spin $\langle S_{\mathbb{I}}^z\rangle$ 
by the surrounding $\pi$ electrons 
for the doped case is 
as large as $\langle S_{\mathbb{I}}^z \rangle \sim 0.4$ (see Fig.~\ref{fig:dmrgn1000_3}), which  
is slightly smaller than that for the undoped case shown in Fig.~\ref{fig:dmrgn1166_3}. 
Interestingly, this tendency has also been observed in the experiments~\cite{nair2}. 

On the other hand, the noticable difference is found in the weak coupling phase. 
As shown in Fig.~\ref{fig:dmrgn1000}, 
$\langle{n}_{i\sigma}\rangle \sim 2/3$ and $\langle{n}_{i\pi}\rangle \sim 1$ for $i\in\mathbb{I}$ in the weak coupling phase, 
and thus the total number $N_{\mathbb{I}}$ of 
electrons in $\mathbb{I}$ is $\langle N_{\mathbb{I}}\rangle\sim5$, which is  
less than that for the undoped case (see Fig.~\ref{fig:dmrgn1166}) and is also less than 6 assumed in Sec.~\ref{sec:multiplet}. 
This is probably the reason 
why the expectation value of total spin squared $ {\bm S}_{\mathbb{I}} \cdot {\bm S}_{\mathbb{I}} $
is smaller than $2$, but $\langle {\bm S}_{\mathbb{I}} \cdot {\bm S}_{\mathbb{I}} \rangle\sim 1.2$, 
corresponding to $\bar{S}_{\mathbb{I}}\sim0.7$,  
in the weak coupling phase. 
However, as in the undoped case, 
the local magnetic moment in the weak coupling phase vanishes due to the coupling 
with the surrounding $\pi$ electron system, as shown in Fig.~\ref{fig:dmrgn1000_3}. 
Therefore, we can conclude that the weak coupling phase is always nonmagnetic
for the undoped and doped cases.

\subsection{\label{sec:spin}Asymptotic behavior of spin correlation function}

Let us now examine the spin correlation function 
$\langle {\bm S}_{\alpha} \cdot \tilde{\bm S}_{lc} \rangle$ 
between the impurity sites and the conduction sites in the Q1D model $\mathcal{H}_{\rm Q1D}$. 
Here, ${\bm S}_{\alpha}$ ($\alpha=\sigma,\pi$) is the total spin operator of the $\alpha$ orbitals around the vacancy, i.e., 
\begin{equation}
{\bm S}_{\sigma} = \sum_{i\in \mathbb{I}} {\bm S}_{i\sigma}
\end{equation}
and
\begin{equation}
{\bm S}_{\pi} = \sum_{i\in \mathbb{I}} {\bm S}_{i\pi},
\end{equation}
and $\tilde{\bm S}_{lc} = (\tilde{S}^{x}_{lc}, \tilde{S}^{y}_{lc}, \tilde{S}^{z}_{lc})$ for $c=a,e_1,e_2$
is the local spin operator of the conduction sites in the Q1D model $\mathcal{H}_{\rm Q1D}$, i.e.,  
\begin{equation}
\tilde{S}_{lc}^{\mu} = \frac{1}{2} \sum_{s,s^{\prime}=\uparrow,\downarrow} 
\tilde{f}_{lcs}^{\dagger} \left[ \mbox{\boldmath{$\sigma$}}^{\mu} \right]_{ss^{\prime}} 
\tilde{f}_{lcs^{\prime}}
\label{eq:cond:spin}
\end{equation}
for $l\ge 2$ [see Fig.~\ref{fig:method}(c)]. Note that 
$\sum_{c=a,e_1,e_2} \tilde{\bm S}_{1c} = {\bm S}_{\pi}$ if we use Eq.~(\ref{eq:cond:spin}) for $l=1$
because the orbitals at $l=1$ correspond to the $\pi$ orbitals in $\mathbb{I}$ around the vacancy, 
as shown in Fig.~\ref{fig:method}.
Since the results for the $c=e_1$ and $e_2$ modes are identical, here we show only one of these results and refer to them simply 
as the results for the $c=e$ mode. 
We also note that the spin correlation function $\langle {\bm S}_{\alpha} \cdot \tilde{\bm S}_{lc} \rangle$ is plotted as a function of 
the distance $d$ between the impurity site and the $l$th conduction site given by $l-1$ in the Q1D model.

\begin{figure}[htbp]
\begin{center}
\includegraphics[width=\hsize]{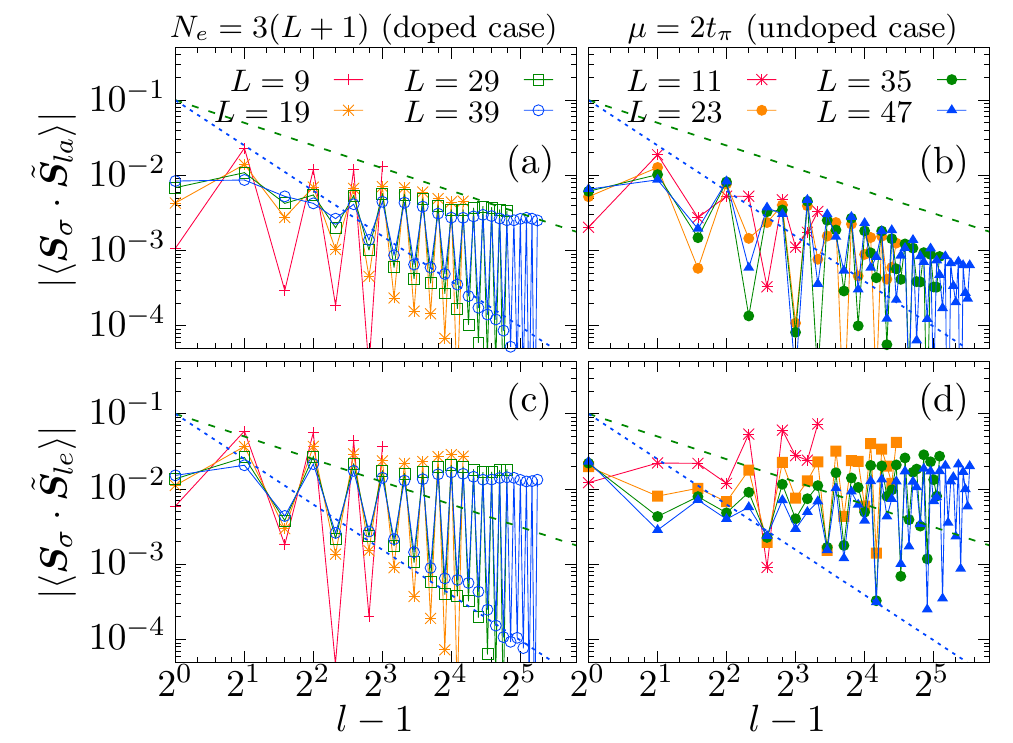}
\caption{
Log-log plots of the absolute values of the spin correlation function $\langle {\bm S}_{\sigma} \cdot \tilde{\bm S}_{lc} \rangle$ 
with the $c=a$ and $e$ modes for the doped case (left panels) and the undoped case (right panels) with various values of 
$L$ in the Q1D model $\mathcal{H}_{\rm Q1D}$. 
The other parameters used are $U=3t$, $J=0.1U$, and $t = 6 t_{\sigma} = 12 t_{\pi}$.
Green dashed line (blue dotted line) represents a function proportional to $(l-1)^{-1}$ [$(l-1)^{-2}$], for comparison. 
}
\label{fig:spinspin}
\end{center}
\end{figure}

Figure~\ref{fig:spinspin} shows the spin correlation function $\langle {\bm S}_{\sigma} \cdot \tilde{\bm S}_{lc} \rangle$  
calculated for the ground states of the doped and undoped cases in the strong coupling phase. 
We can find in Fig.~\ref{fig:spinspin} that 
$\langle {\bm S}_{\sigma} \cdot \tilde{\bm S}_{lc} \rangle$ decays as or even
slower than $d^{-1}$ for both $c=a$ and $e$ modes.
Note also that such behavior does not depend qualitatively on 
the carrier doping, and thus the spin structure of 
the $\sigma$ orbitals is robust against the carrier doping. 
However, the overall values 
of the spin correlation function for the $e$ mode is approximately $10$ times larger 
than those for the $a$ mode,
suggesting that the local spin ${\bm S}_{\sigma}$ of the $\sigma$ orbitals around the vacancy 
is mostly coupled to the surrounding $\pi$ orbitals with the $c=e$ mode.

As an effective model of the Anderson model $\mathcal{H}_{\rm AM}$, 
we consider in Appendix~\ref{app:refspin} a two-orbital Anderson model $\mathcal{H}^{({\text {II}})}$ 
described in Eq.~(\ref{eq:refmodel:II}),  
referred to as ``model II" in Appendix~\ref{app:refspin}.  
As schematically shown in Fig.~\ref{fig:repmodel}(b), the impurity sites in this model are composed of 
one of the lattice sites in the honeycomb lattice that is replaced with a Hubbard site (denoted as impurity site B) 
and an additional Hubbard site (denoted as impurity site A) attached on top of the impurity site B. These impurity sites 
contains the inter-orbital Coulomb interactions with no hopping and are coupled to the conduction sites through the impurity site B 
with a diverging hybridization function. 
Therefore, corresponding the impurity sites A and B in $\mathcal{H}^{({\text {II}})}$ to 
the $\sigma$ and $\pi$ orbitals in $\mathbb{I}$ for $\mathcal{H}_{\rm AM}$, respectively, 
we can consider this model $\mathcal{H}^{({\text {II}})}$ as a simple version of the Anderson model 
$\mathcal{H}_{\rm AM}$. 

As shown in Figs.~\ref{fig:repspin2}(a) and \ref{fig:repspin2}(b) for the doped and undoped cases, respectively, 
we find that the spin correlation function between 
the impurity site A and the conduction sites in the Q1D representation of 
$\mathcal{H}^{({\text {II}})}$ [see Fig.~\ref{fig:repmodel}(d)]
decays approximately as $d^{-1}$, 
which resemble the results shown in Figs.~\ref{fig:spinspin}(a) and \ref{fig:spinspin}(b). 
In the strong coupling phase of this two-orbital Anderson model $\mathcal{H}^{({\text {II}})}$, the Hund's coupling 
favors the formation of the local spin-1 state at the impurity sites, while the effect of 
the surrounding electrons in the conduction sites partially screen this spin-1 state, 
thus leading to the emergence of the spin-1/2 free magnetic moment. 
The partial screening in this model is robust even for the doped case because of the large hybridization.
These features are indeed similar to the underlying picture obtained for the Anderson model $\mathcal{H}_{\rm AM}$ 
in the previous sections. 
Namely, the local spin $\bar{S}_{\mathbb{I}} \sim 1$ around the vacancy 
is partially screened by the surrounding $\pi$ electrons in the conduction sites and the residual 
free magnetic moment is robust against the doping.

\begin{figure}[htbp]
\begin{center}
\includegraphics[width=\hsize]{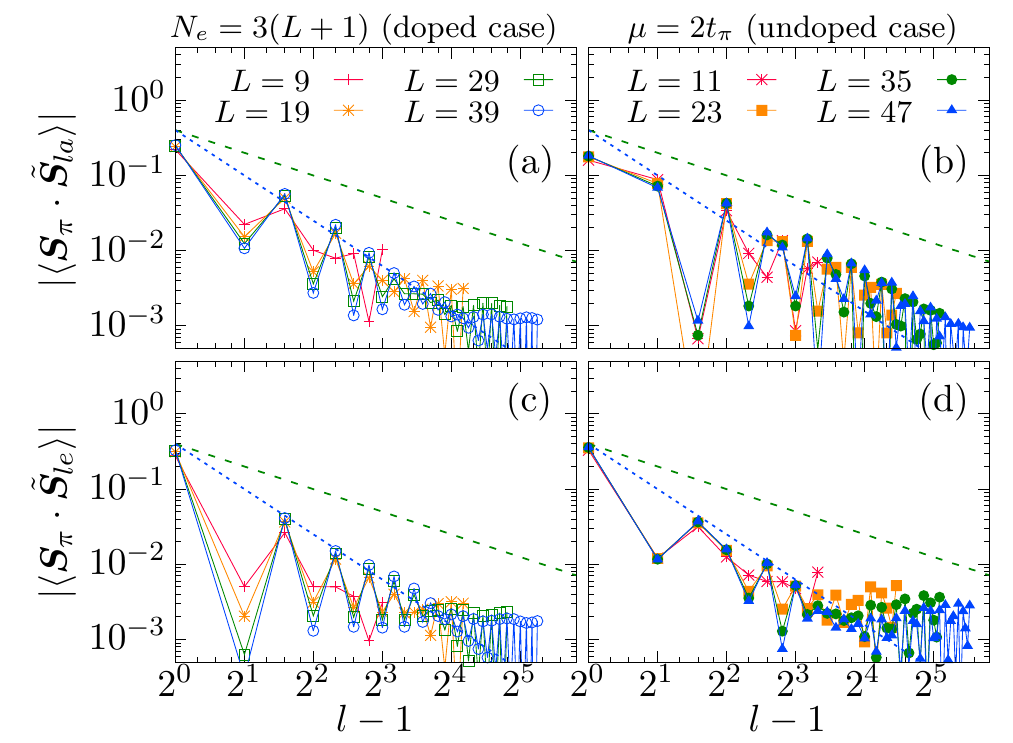}
\caption{
Same as Fig.~\ref{fig:spinspin} except for the spin correlation function $\langle {\bm S}_{\pi} \cdot \tilde{\bm S}_{lc} \rangle$ 
with the $c=a$ and $e$ modes. }
\label{fig:spinspin2}
\end{center}
\end{figure}

Figure~\ref{fig:spinspin2} shows the spin correlation function $\langle {\bm S}_{\pi} \cdot \tilde{\bm S}_{lc} \rangle$ 
for the ground states of the doped and undoped cases of the Q1D model $\mathcal{H}_{\rm Q1D}$ in the strong coupling phase. 
We find in Figs.~\ref{fig:spinspin2}(c) and \ref{fig:spinspin2}(d) 
that $\langle {\bm S}_{\pi} \cdot \tilde{\bm S}_{le} \rangle$ 
decays approximately as $d^{-2}$ for both undoped and doped cases. 
The spin correlation function $\langle {\bm S}_{\pi} \cdot \tilde{\bm S}_{la} \rangle$ 
for the doped case in Fig.~\ref{fig:spinspin2}(a) also exhibits the $d^{-2}$ decay.
However, $\langle {\bm S}_{\pi} \cdot \tilde{\bm S}_{la} \rangle$ 
for the undoped case in Fig.~\ref{fig:spinspin2}(b) is different and is instead somewhat enhanced as 
compared with the other cases: it deviates from the $d^{-2}$ decay 
but rather approaches to the $d^{-1}$ decay. 

As described in Appendix~\ref{app:refspin}, the $d^{-2}$ decay of the spin correlation function implies 
the absence of the free magnetic moment. 
We find in Figs.~\ref{fig:repspin2}(c) and \ref{fig:repspin2}(d) that the spin correlation function 
between the impurity site B and the conduction sites 
decays as $d^{-2}$ for both doped and undoped cases of the two-orbital Anderson model $\mathcal{H}^{({\text {II}})}$. 
In this model, the screening mainly occurs at the impurity site B among the two impurity sites via the diverging hybridization function. 
This is indeed similar to the features found here for the Anderson model $\mathcal{H}_{\rm AM}$ because 
the local spin $\bar{S}_{\mathbb{I}} \sim 1$ around the vacancy is partially screened by the surrounding $\pi$ electrons 
in the conduction band
via the diverging hybridization function $\tilde{\Gamma}_e(\omega)$, leading to the $d^{-2}$ decay 
of the spin correlation function $\langle {\bm S}_{\pi} \cdot \tilde{\bm S}_{le} \rangle$.

In contrast, the spin correlation function $\langle {\bm S}_{\pi} \cdot \tilde{\bm S}_{la} \rangle$ 
in the largest system for the undoped case shown in Fig.~\ref{fig:spinspin2}(b) decays as $d^{-1.43}$, 
certainly slower than $d^{-2}$, 
which is a hallmark of the presence of the free magnetic moment
(see the discussion in Appendix~\ref{app:refspin}).  
This can be 
attributed to the fact that the hybridization function $\tilde{\Gamma}_a(\omega)$ exhibits 
the pseudogap structure at the Fermi level for the undoped case, as shown in Fig.~\ref{fig:dhyb}. 
It is known that the Kondo screening does not occur in the undoped pseudogap Anderson model because 
the density of states vanishes at the Fermi energy~\cite{bulla,fritz,shirakawa2,gonzalez-buxton,vojta}. 
In this case, there are two possible phases that arises when the carriers are introduced by varying the chemical potential away from 
the pseudogap position. 
One is the Kondo screening phase simply because the density of state becomes finite 
when the chemical potential is shifted from the pseudogap position~\cite{vojta}.
The other is the asymmetric strong coupling phase, in which 
the electronic state at the impurity site becomes either 
empty or doubly occupied~\cite{vojta}.
Both in the Kondo screening and asymmetric strong coupling phases, 
the magnetic moment vanishes. In either case, therefore, 
the free magnetic moment is sensitive against the carrier doping  
(see Appendix~\ref{app:refspin} for the numerical study of a simple model for the pseudogap Anderson system).
This might explain why the residual free magnetic moment around the vacancy is slightly reduced when the carriers are 
introduced in the strong coupling phase of the Anderson model $\mathcal{H}_{\rm AM}$ 
(see Figs.~\ref{fig:dmrgn1166_3} and \ref{fig:dmrgn1000_3}).

\section{\label{sec:discussion}Summary and Discussion}

We have studied the ground state properties of the effective Anderson model for a single vacancy in graphene.  
The Anderson model considered here is composed of the three dangling $\sigma$ orbitals as well as the three $\pi$ orbitals 
of carbon atoms around the vacancy,  
treated as the impurity sites, which are coupled to the noninteracting conduction electrons in the surrounding $\pi$ orbitals of 
carbon atoms forming the honeycomb lattice with the single vacancy. 
Employing the block Lanczos DMRG method, 
we have found that the ground state in the reasonable parameter region for graphene, belonging to the strong coupling phase,  
shows the emergence of the free magnetic moment with its spin as large as $\sim1/2$ around the vacancy. 
Our systematic analysis of the projected weight onto the local multiplet states has shown  
that the local spin ${S}_{\mathbb{I}}=1$ multiplet with the doubly degenerate $\mathcal{E}$ representation 
of the $\mathcal{C}_3$ point group is dominant in the ground state, and thus 
the residual free magnetic moment is attributed to 
the partial screening of this ${S}_{\mathbb{I}}=1$ multiplet by the sorrounding $\pi$ electrons in the conduction band.
We have also found that the emergent local free magnetic moment is robust against the hole doping. 
The calculations of the spin correlation function have shown that
the screening involves mostly the local $\pi$ orbitals around the vacancy and 
thus the residual free magnetic moment is composed mainly of the local $\sigma$ orbitals. 
Because these $\sigma$ orbitals are decoupled with no happing terms to the surrounding $\pi$ orbital system, 
the emergent local free magnetic moment is not sensitive to any change of the surrounding $\pi$ orbital system 
such as the carrier doping. However, our calculations of the spin correlation function 
have also shown the precursor indicating another contribution to the emergent local 
free magnetic moment in the undoped case.
This contribution arises due to the pseudogap structure 
of the hybridization function $\tilde{\Gamma}_a(\omega)$ in the $\pi$ orbital system
and thus sensitive to the carrier doping.
The presence of these two contributions to the emergent local free magnetic moment
fits nicely the experimental observations~\cite{nair2,yzhang}
and their scenario proposed in Ref.~\cite{nair2}.

It should be noted that the dominant contribution to the local multiplet structure around the vacancy, 
having  the local spin 1 and  the $\mathcal{E}$ irreducible representation, 
is compatible with the occurrence of the in-plane Jahn-Teller distortion 
found in the previous {\it ab-initio} DFT calculations~\cite{paz,padmanabhan}, 
in a sense that the Jahn-Teller distortion occurs 
through the coupling with the in-plane $e$ vibration mode~\cite{casartelli},  
similar to the so-called $\mathbb{E} \otimes e$ problem~\cite{bersuker}.
However, our results imply that 
the Jahn-Teller distortion is not necessarily required to form 
the local spin triplets and the residual free magnetic moment as large as spin $S=1/2$.

In contrast, the out-of-plane distortion generates a finite hybridization between the neighboring
$\sigma$ and $\pi$ orbitals. As a consequence, a nonmagnetic ground state can be realized 
because the residual free magnetic moment composed mainly of the $\sigma$ orbitals around the vacancy 
would be screened by the surrounding $\pi$ electrons via 
this finite hybridization~\cite{kanao,shirakawa1}. 
This conjecture is also in accordance with the previous {\it ab-initio} DFT results~\cite{paz,padmanabhan}, 
showing that the out-of-plane distortion gives rise to a nonmagnetic ground state. 
This may bring an interesting functionality by controlling the out-of-plane strain 
field suggested by the {\it ab-initio} DFT calculation~\cite{dharma-wardana}.

The scanning tunneling spectroscopy measurements of a single vacancy in graphene 
have revealed two peaks in the tunneling spectrum as a function of bias voltage 
that are originated from the spin-polarized states around 
the vacancy~\cite{yzhang}. They also found that these two peaks are quite robust 
against the carrier doping into graphene~\cite{nair2,yzhang}. This is in sharp contrast 
to the magnetism in graphene induced by hydrogen absorption~\cite{gonzalez-herrerd}, 
where the magnetic moment is sensitive to the carrier doping. 
This is also supported by the theoretical analysis on the carrier doping 
in the pseudogap Kondo model~\cite{vojta}.
Our results in this paper as well as 
the previous calculations of effective models for a single adatom in graphene~\cite{shirakawa1,shirakawa2} 
suggest that the magnetism observed in these two systems is caused by different mechanisms. 
Here, we have focused on static physical quantities because 
the calculation of dynamical quantities is rather demanding.  
However, it is highly valuable to examine 
the dynamical quantities such as the local density of states and the response to an external field including a gauge field, 
which can shed light on the relation between the local multiplet structures and their experimental observation. 
These studies are left for the future.

\section*{Acknowledgements} 
The authors are grateful to Beom Hyn Kim for fruitful discussions. 
A part of the numerical simulations has been performed
using the HOKUSAI supercomputer at RIKEN
(Project ID: Q21532) and also supercomputer Fugaku installed in RIKEN R-CCS.
This work is supported by 
Grant-in-Aid for Scientific Research (B) (No.~JP18H01183) and 
Grant-in-Aid for Scientific Research (A) (No.~JP21H04446) from MEXT, Japan. 
This work is also supported in part by the COE research grant in computational science from 
Hyogo Prefecture and Kobe City through Foundation for Computational Science.

\appendix

\section{\label{app:localmultiplet}Local multiplet states in the strong coupling limit and projected weight}

In this appendix, we shall construct explicitly the eigenstates of the local part $\mathcal{H}_{\mathbb{I}}$ 
of the Anderson model $\mathcal{H}_{\rm AM}$ in the strong coupling limit, assuming that the number of electrons is six. 
We also provide the definition of the projected weight $P( \mathbb{C}, S_{\mathbb{I}} )$ with $\mathbb{C}=\mathbb{A}, \mathbb{E}$. 

\subsection{Local multiplet states}

As described in Sec.~\ref{sec:multiplet}, the low-lying states in the strong coupling limit 
form the local spin one at each carbon site that is occupied by two electrons. Therefore, the local Hilbert space 
at site $i$~($i=1,2,3$)~$\in\mathbb{I}$ [see Fig.~\ref{fig:model}(c)] can be spanned by the following bases: 
\begin{align}
& \vert 1 \rangle_i  = c_{i\sigma\uparrow}^{\dagger} c_{i\pi \uparrow}^{\dagger} \vert 0 \rangle, \nonumber \\
& \vert 0 \rangle_i  = \frac{1}{\sqrt{2}} 
( c_{i\sigma\uparrow}^{\dagger} c_{i\pi\downarrow}^{\dagger} + 
c_{i\sigma\downarrow}^{\dagger} c_{i\pi\uparrow}^{\dagger} )
\vert 0 \rangle, \label{eq:spinonbasis} \\
& \vert -1 \rangle_i = c_{i\sigma\downarrow}^{\dagger} c_{i\pi \downarrow}^{\dagger} \vert 0 \rangle, \nonumber 
\end{align}
where $|S_i^z\rangle_i$ with $S_i^z=-1,0,1$ is the simultaneous eigenstate of the total spin operator ${\bm S}_{i} $ 
defined in Eq.~(\ref{eq:localspin}) and 
the $z$-component of ${\bm S}_{i} $ at site $i$ with its eigenvalues 1 and $S_i^z$, respectively, 
and $c_{i\alpha s}^{\dagger}$ is the 
creation operator of an electron at site $i$ with spin $s$ ($=\uparrow,\downarrow$)
and orbital $\alpha$ (=$\sigma$, $\pi$).

The effective low-energy Hamiltonian is then given by the three-site spin-one antiferromagnetic Heisenberg model,
\begin{equation}
\mathcal{H}_{\rm S} = J_{\rm ex} \sum_{\scriptsize\llangle i,j \rrangle_{\mathbb{I}} } 
{\bm S}_i \cdot {\bm S}_j,
\end{equation}
where $J_{\rm ex}$ is an effective exchange interaction, 
$\llangle i, j \rrangle_{\mathbb{I}}$ denotes a pair of neighboring sites $i$ and $j$ around the vacancy, 
and ${\bm S}_i$ is the spin-1 operator at site $i \in \mathbb{I}$ with its eigenstates being given  
by Eq.~(\ref{eq:spinonbasis}).
For the three-site case, this Hamiltonian is particularly simplified by 
introducing the total spin operator 
${\bm S}_{\mathbb{I}}=\sum_{i\in\mathbb{I}}{\bm S}_i=(S_{\mathbb{I}}^x,S_{\mathbb{I}}^y,S_{\mathbb{I}}^z)$ as  
\begin{equation}
\mathcal{H}_{\rm S} = \frac{J_{\rm ex}}{2} \left( {\bm S}_{\mathbb{I}} \cdot {\bm S}_{\mathbb{I}} - 6 \right).
\end{equation}
Therefore, we can readily find that the eigenvalues of $\mathcal{H}_{\rm S}$ are 
determined solely by the quantum number of ${\bm S}_{\mathbb{I}}$, 
denoted as $S_{\mathbb{I}}$. The eigenvalues $E(S_{\mathbb{I}})$ 
of $\mathcal{H}_{\rm S}$ are thus given as  
\begin{equation}
\begin{split}
& E(S_{\mathbb{I}} = 0 ) = - 3 J_{\rm ex}, \\
& E(S_{\mathbb{I}} = 1 ) = - 2 J_{\rm ex}, \\
& E(S_{\mathbb{I}} = 2 ) = 0, \\
& E(S_{\mathbb{I}} = 3 ) = 3 J_{\rm ex}. \\
\end{split}
\label{eq:energyeigenvalues}
\end{equation}

Because of the $120^{\circ}$ rotational symmetry around the center of the cluster, 
the eigenstates of $\mathcal{H}_{\rm S}$ are also characterized by the irreducible representations of the 
$\mathcal{C}_3$ point group. 
Therefore, we can denote the energy eigenstate as $\vert C, S_{\mathbb{I}}, M_{\mathbb{I}} \rangle$, 
where $C = A $ ($C=E_k $ with $k=1,2$) indicates 
the symmetric (doubly degenerate) irreducible representation $\mathcal{A}$ ($\mathcal{E}$) of the $\mathcal{C}_3$ point group, and 
$M_{\mathbb{I}}$ denotes the quantum number of the $z$-component $S^z_{\mathbb{I}}$ of the total spin operator ${\bm S}_i$. 
The lowest eigenstate is $(\mathbb{A},0)$ [for notation, see Sec.~\ref{sec:multiplet} and Fig.~\ref{fig:hloc}(d)] and is 
given by
\begin{align}
\vert A, 0, 0 \rangle & 
= 
\frac{1}{\sqrt{6}} 
\left[ 
\vert 0 \rangle_1 \vert 1 \rangle_2 \vert -1 \rangle_3 - 
\vert 0 \rangle_1 \vert -1 \rangle_2 \vert 1 \rangle_3 \right]
\nonumber \\
& + 
\frac{1}{\sqrt{6}} 
\left[ 
\vert -1 \rangle_1 \vert 0 \rangle_2 \vert 1 \rangle_3 -
\vert 1 \rangle_1 \vert 0 \rangle_2 \vert -1 \rangle_3 \right]
\nonumber \\
& + 
\frac{1}{\sqrt{6}}
\left[
\vert 1 \rangle_1 \vert -1 \rangle_2 \vert 0 \rangle_3 -
\vert -1 \rangle_1 \vert 1 \rangle_2 \vert 0 \rangle_3 \right],
\label{eq:multipleta0s0}
\end{align}
where $\vert S_i^z \rangle_i$ with $S_i^z=-1,0,1$ and $i=1,2,3$ is given in Eq.~(\ref{eq:spinonbasis}).

The second lowest eigenstates are those with $S_{\mathbb{I}} = 1$, 
which are futher classified by the irreducible representation $C$, i.e., 
$(\mathbb{A},1)$ and $(\mathbb{E},1)$. 
The eigenstate for $(\mathbb{A},1)$ with $M_{\mathbb{I}}^z = 0$ 
is given by 
\begin{align}
\vert A, 1, 0 \rangle & = 
\frac{1}{\sqrt{15}} 
\left[ 
\vert 0 \rangle_1 \vert 1 \rangle_2 \vert -1 \rangle_3 + 
\vert 0 \rangle_1 \vert -1 \rangle_2 \vert 1 \rangle_3
\right] \nonumber \\
& + 
\frac{1}{\sqrt{15}} 
\left[ 
\vert -1 \rangle_1 \vert 0 \rangle_2 \vert 1 \rangle_3 + 
\vert 1 \rangle_1 \vert 0 \rangle_2 \vert -1 \rangle_3
\right] \nonumber \\
& + 
\frac{1}{\sqrt{15}} 
\left[ 
\vert 1 \rangle_1 \vert -1 \rangle_2 \vert 0 \rangle_3 + 
\vert -1 \rangle_1 \vert 1 \rangle_2 \vert 0 \rangle_3
\right] \nonumber \\
& -
\frac{\sqrt{3}}{\sqrt{5}} 
\vert 0 \rangle_1 \vert 0 \rangle_2 \vert 0 \rangle_3. 
\label{eq:multipleta0s1}
\end{align}
The eigenstates with different $M_{\mathbb{I}}$ ($=\pm 1$) 
are obtained by multiplying $S_{\mathbb{I}}^\pm=S_{\mathbb{I}}^x\pm{\rm i}S_{\mathbb{I}}^y$ operators. 
Similarly, the other two eigenstates for $(\mathbb{E},1)$ 
with $M_{\mathbb{I}}=0$ are given by  
\begin{align}
\vert E_1, 1, 0 \rangle & 
= 
\frac{2}{2\sqrt{3}} 
\left[ 
\vert 0 \rangle_1 \vert 1 \rangle_2 \vert -1 \rangle_3 +
\vert 0 \rangle_1 \vert -1 \rangle_2 \vert 1 \rangle_3
\right] \nonumber \\
& 
-
\frac{1}{2\sqrt{3}} 
\left[ 
\vert -1 \rangle_1 \vert 0 \rangle_2 \vert 1 \rangle_3 +
\vert 1 \rangle_1 \vert 0 \rangle_2 \vert -1 \rangle_3
\right] \nonumber \\
&
-
\frac{1}{2\sqrt{3}} 
\left[ 
\vert 1 \rangle_1 \vert -1 \rangle_2 \vert 0 \rangle_3 +
\vert -1 \rangle_1 \vert 1 \rangle_2 \vert 0 \rangle_3
\right]
\label{eq:multiplete1s1}
\end{align}
and 
\begin{align}
\vert E_2, 1, 0 \rangle & = 
\frac{1}{2} 
\left[ 
\vert -1 \rangle_1 \vert 0 \rangle_2 \vert 1 \rangle_3 +
\vert 1 \rangle_1 \vert 0 \rangle_2 \vert -1 \rangle_3
\right] \nonumber \\
&
-
\frac{1}{2} 
\left[ 
\vert 1 \rangle_1 \vert -1 \rangle_2 \vert 0 \rangle_3 +
\vert -1 \rangle_1 \vert 1 \rangle_2 \vert 0 \rangle_3
\right]. 
\label{eq:multiplete2s1}
\end{align}

The third eigenstates are those for $(\mathbb{E},2)$. 
The eigenstates with $M_{\mathbb{I}}=0$ are given by  
\begin{align}
\vert E_1, 2, 0 \rangle & = 
\frac{2}{2\sqrt{3}}
\left[ 
\vert 0 \rangle_1 \vert 1 \rangle_2 \vert -1 \rangle_3 -
\vert 0 \rangle_1 \vert -1 \rangle_2 \vert 1 \rangle_3
\right] \nonumber \\
& -
\frac{1}{2\sqrt{3}} 
\left[ 
\vert -1 \rangle_1 \vert 0 \rangle_2 \vert 1 \rangle_3 -
\vert 1 \rangle_1 \vert 0 \rangle_2 \vert -1 \rangle_3
\right] \nonumber \\
& -
\frac{1}{2\sqrt{3}} 
\left[ 
\vert 1 \rangle_1 \vert -1 \rangle_2 \vert 0 \rangle_3 -
\vert -1 \rangle_1 \vert 1 \rangle_2 \vert 0 \rangle_3
\right] 
\label{eq:multiplete1s2}
\end{align}
and 
\begin{align}
\vert E_2, 2, 0 \rangle & =
\frac{1}{2} 
\left[ 
\vert -1 \rangle_1 \vert 0 \rangle_2 \vert 1 \rangle_3 -
\vert 1 \rangle_1 \vert 0 \rangle_2 \vert -1 \rangle_3
\right] \nonumber \\
&
-
\frac{1}{2} 
\left[ 
\vert 1 \rangle_1 \vert -1 \rangle_2 \vert 0 \rangle_3 -
\vert -1 \rangle_1 \vert 1 \rangle_2 \vert 0 \rangle_3
\right] .
\label{eq:multiplete2s2}
\end{align}
Finally, the fourth eigenstates are those for $(\mathbb{A},3)$. 
The eigenstates with $M_{\mathbb{I}}=0$ is given by 
\begin{align}
\vert A, 3, 0 \rangle & = 
\frac{1}{\sqrt{10}} 
\left[
\vert 0 \rangle_1 \vert 1 \rangle_2 \vert -1 \rangle_3 +
\vert 0 \rangle_1 \vert -1 \rangle_2 \vert 1 \rangle_3
\right]
\nonumber \\
& + 
\frac{1}{\sqrt{10}} 
\left[ 
\vert -1 \rangle_1 \vert 0 \rangle_2 \vert 1 \rangle_3 +
\vert 1 \rangle_1 \vert 0 \rangle_2 \vert -1 \rangle_3 \right]
\nonumber \\
& + 
\frac{1}{\sqrt{10}}
\left[
\vert 1 \rangle_1 \vert -1 \rangle_2 \vert 0 \rangle_3 +
\vert -1 \rangle_1 \vert 1 \rangle_2 \vert 0 \rangle_3 \right]
\nonumber \\
& +
\frac{\sqrt{2}}{\sqrt{10}} 
\vert 0 \rangle_1 \vert 0 \rangle_2 \vert 0 \rangle_3. 
\label{eq:multipleta0s3}
\end{align}

\subsection{Projected weight}

Let $\mathcal{Q}^{\dagger}_{C S_{\mathbb{I}}M_{\mathbb{I}}}$ be an operator that generates the eigenstate 
$\vert C, S_{\mathbb{I}}, M_{\mathbb{I}} \rangle_\mathbb{I}$ from the vacuum state $|0\rangle_{\mathbb{I}}$ 
in the Hilbert space of the region $\mathbb{I}$, i.e., 
\begin{align}
\vert C, S_{\mathbb{I}}, M_{\mathbb{I}} \rangle_\mathbb{I} 
= \mathcal{Q}^{\dagger}_{C S_{\mathbb{I}} M_{\mathbb{I}}} \vert 0 \rangle_\mathbb{I},
\label{eq:pop1}
\end{align}
where $\vert C, S_{\mathbb{I}}, M_{\mathbb{I}} \rangle_\mathbb{I}$ 
($C = A$, $E_1$, $E_2$) 
is given in Eqs.~(\ref{eq:multipleta0s0})-(\ref{eq:multipleta0s3}). 
Replacing all $\vert S_1^z \rangle_1 \vert S_2^z \rangle_2 \vert S_3^z \rangle_3$ 
in $\vert C, S_{\mathbb{I}}, M_{\mathbb{I}} \rangle_\mathbb{I}$ 
with the electron creation operators as in Eqs.~(\ref{eq:spinonbasis}), 
$\mathcal{Q}^{\dagger}_{C S_{\mathbb{I}} M_{\mathbb{I}}}$ are 
now represented as the linear combination of products of six electron creation operators. 
We then introduce operators $\mathcal{N}_0$ and $\mathcal{P}_{C S_{\mathbb{I}} M_{\mathbb{I}}}^{\dagger}$ 
defined respectively by 
\begin{equation}
\mathcal{N}_0 = \prod_{i \in \mathbb{I}} \prod_{\alpha = \sigma, \pi} \prod_{s = \uparrow,\downarrow} 
c_{i \alpha s} c_{i \alpha s}^{\dagger} 
\end{equation}
and
\begin{equation}
\mathcal{P}^{\dagger}_{C S_{\mathbb{I}} M_{\mathbb{I}}} = 
\mathcal{Q}^{\dagger}_{C S_{\mathbb{I}} M_{\mathbb{I}}} \mathcal{N}_0. 
\end{equation}
We can readily find that $\mathcal{N}_0$ is a projection operator onto the vacuum state 
in the region $\mathbb{I}$.
Note also that $\mathcal{P}^{\dagger}_{C S_{\mathbb{I}} M_{\mathbb{I}}}$ 
can be represented as the linear combination of products of eighteen electron creation and annihilation operators.

Let $\vert \Phi_0 \rangle$ be the ground state of the whole system of the Anderson model $\mathcal{H}_{\rm AM}$. 
In general, the ground state $\vert \Phi_0 \rangle$ can be written as 
\begin{equation}
\vert \Phi_0 \rangle = \sum_{n,l} \phi_{nl} |\psi_n\rangle_{\mathbb{I}} |\tilde{\psi}_l\rangle_{\mathbb{L}},
\end{equation}
where $|\psi_n\rangle_{\mathbb{I}}$ and $|\tilde{\psi}_l\rangle_{\mathbb{L}}$ are orthonormalized basis states in 
the regions $\mathbb{I}$ and $\mathbb{L}$, respectively, and we assume that the ground state $\vert \Phi_0 \rangle$ is normalized. 
The reduced density matrix operator $\rho_{\mathbb{I}}$ in the region $\mathbb{I}$ is then given by
\begin{equation}
\rho_{\mathbb{I}} = {\rm Tr}_{\mathbb{L}} \vert \Phi_0 \rangle \langle \Phi_0 \vert,
\end{equation}
where ${\rm Tr}_{\mathbb{L}}\cdots$ denotes the trace over all the basis states in the region $\mathbb{L}$. 
It is now easy to show that $_\mathbb{I}\langle\psi_n| \rho_{\mathbb{I}} |\psi_n\rangle_{\mathbb{I}} = \sum_l |\phi_{nl}|^2$. 
The projected weight $P( \mathbb{C}, S_{\mathbb{I}} )$ discussed in Secs.~\ref{sec:undoped} and \ref{sec:doped}
are defined by  
\begin{equation}
\begin{split}
P( \mathbb{A}, S_{\mathbb{I}} )
& 
= \sum_{M_{\mathbb{I}}=-S_{\mathbb{I}}}^{S_{\mathbb{I}}} 
\langle A, S_{\mathbb{I}}, M_{\mathbb{I}} \vert 
\rho_{\mathbb{I}} 
| A, S_{\mathbb{I}}, M_{\mathbb{I}} \rangle, \\
P( \mathbb{E}, S_{\mathbb{I}} )
& 
= \sum_{k=1}^2 \sum_{M_{\mathbb{I}}=-S_{\mathbb{I}}}^{S_{\mathbb{I}}} 
\langle E_k, S_{\mathbb{I}}, M_{\mathbb{I}} \vert 
\rho_{\mathbb{I}}
\vert E_k, S_{\mathbb{I}}, M_{\mathbb{I}} \rangle. 
\end{split}
\end{equation}

To reduce the numerical complexity of the DMRG calculations, we evaluate the following equivalent quantities   
\begin{equation}
\begin{split}
P( \mathbb{A}, S_{\mathbb{I}} )
& 
= \sum_{M_{\mathbb{I}}=-S_{\mathbb{I}}}^{S_{\mathbb{I}}} 
\langle \Phi_0 \vert 
\mathcal{P}_{A S_{\mathbb{I}} M_{\mathbb{I}}}^{\dagger} 
\mathcal{P}_{A S_{\mathbb{I}} M_{\mathbb{I}}}
\vert \Phi_0 \rangle, \\
P( \mathbb{E}, S_{\mathbb{I}} )
& 
= \sum_{k=1}^2 \sum_{M_{\mathbb{I}}=-S_{\mathbb{I}}}^{S_{\mathbb{I}}} 
\langle \Phi_0 \vert 
\mathcal{P}_{E_k S_{\mathbb{I}} M_{\mathbb{I}}}^{\dagger}
\mathcal{P}_{E_k S_{\mathbb{I}} M_{\mathbb{I}}}
\vert \Phi_0 \rangle 
\end{split}
\label{eq:projectedweight}
\end{equation}
by using the multitarget technique~\cite{hallberg}, for which  
the ground state $\vert \Phi_0 \rangle$ and 
$\mathcal{P}_{C {S}_{\mathbb{I}} M_{\mathbb{I}}} \vert \Phi_0 \rangle$ 
are included as the target states.

\section{\label{app:refspin}Asymptotic behavior of spin correlation function for reference models}

In order to better understand the results for the Anderson model $\mathcal{H}_{\rm AM}$ in Sec.~\ref{sec:spin}, 
here in this appendix we consider two simpler Anderson models (model I and model II) 
as reference systems and provide the results 
of the spin correlation function between the impurity and conduction sites for these models. 

\begin{figure}[htbp]
\includegraphics[width=\hsize]{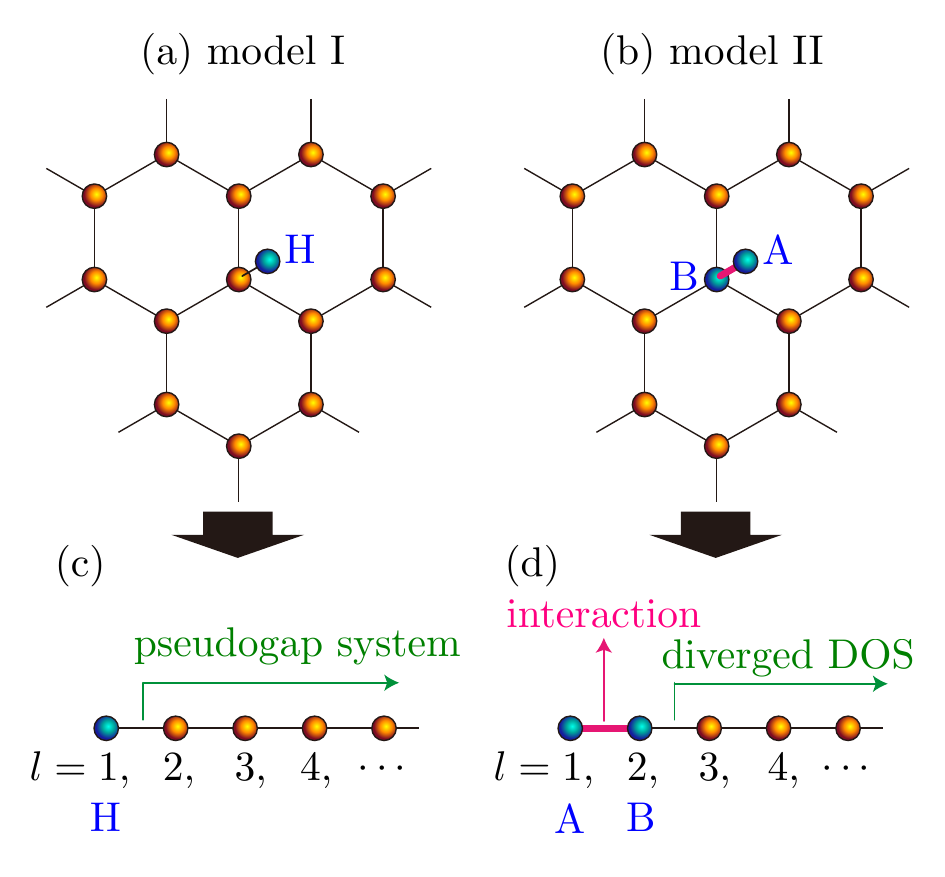}
\caption{
Schematic depictions of (a) model I (a single-impurity Anderson model) and 
(b) model II (a two-impurity Anderson model). 
The impurity sites are indicated with ``H" in (a) and ``A" and ``B" in (b). 
Schematic depictions of the Q1D representation of (c) model I and (d) model II. 
Blue and orange spheres denote the impurity and conduction sites, respectively. 
Black and bold magenta lines indicate lattice bonds connecting two neighboring sites 
via the hopping and two-body interactions, respectively. 
Local density of states at $l=2$ without the impurity site in (c) shows a pseudogap structure,
while local density of states at $l=3$ without the impurity sites in (d) exhibits a diverging structure, 
at Fermi level for the undoped case. 
The impurity site at $l=1$ in (c) is exactly the same as the impurity site $H$ in (a), 
and the impurity sites at $l=1$ and 2 in (d) are exactly the same as the impurity sites A and B, respectively, in (b). 
}
\label{fig:repmodel}
\end{figure}

As shown schematically in Fig.~\ref{fig:repmodel}(a), model I is a single-impurity Anderson model with a single impurity site attached 
to one of the conduction sites in the honeycomb lattice. The Hamiltonian $\mathcal{H}^{(\text{I})} $ of model I is given by  
\begin{align}
\mathcal{H}^{(\text{I})} & = 
U n_{{\rm H} \uparrow} n_{{\rm H} \downarrow} 
- \frac{U}{2} (n_{{\rm H} \uparrow} + n_{{\rm H} \downarrow}), \nonumber \\
& - V \sum_{ s=\uparrow,\downarrow} (c_{{\rm H}  s}^{\dagger} c_{i_0 \pi  s} + c_{i_0 \pi  s}^{\dagger} c_{{\rm H}  s}) \nonumber \\
& - t  \sum_{\langle i, j \rangle} \sum_{ s=\uparrow,\downarrow}
( c_{i \pi  s}^{\dagger} c_{j \pi  s} + c_{j \pi  s}^{\dagger} c_{i \pi  s} ) \nonumber \\
& - t_{\pi}  \sum_{\mbox{\scriptsize{$\llangle$}} i, j \mbox{\scriptsize{$\rrangle$}}} \sum_{ s=\uparrow,\downarrow}
( c_{i \pi  s}^{\dagger} c_{j \pi  s} + c_{j \pi  s}^{\dagger} c_{i \pi  s} ),
\label{eq:refmodel:I}
\end{align}
where $c_{{\rm H} s}$ ($c_{{\rm H} s}^{\dagger}$) is the annihilation (creation) operator 
of an electron at the impurity site H with spin $s\,(=\uparrow,\downarrow)$ and $n_{{\rm H} s} = c_{{\rm H} s}^{\dagger} c_{{\rm H} s}$. 
$c_{i\pi s}$ ($c_{i\pi s}^{\dagger}$) is the annihilation (creation) operator of an electron at conduction site $i$ with 
spin $s\,(=\uparrow,\downarrow)$. 
$i_0$ labels the conduction site that is connected to the impurity site H through the hopping $V$. 
As in the Anderson model $\mathcal{H}_{\rm AM}$, the nearest and next nearest neighboring hoppings $t$ and $t_\pi$, respectively, 
are considered in the last two terms in Eq.~(\ref{eq:refmodel:I}).

As shown schematically in Fig.~\ref{fig:repmodel}(b), model II is a two-impurity (or two-orbital) Anderson model, where 
the impurity sites are composed of one of the lattice sites in the honeycomb lattice (referred to as the impurity site B) and 
an additional impurity site (referred to as the impurity site A) attached on top of the impurity site B through the inter-orbital interactions 
without any hopping. Therefore, only the impurity site B is connected to the conduction sites. 
Comparing the Anderson model $\mathcal{H}_{\rm AM}$ in Eq.~(\ref{eq:am}), the impurity site A (B) mimics the $\sigma$ ($\pi$) orbitals 
in the impurity sites (i.e., in the region $\mathbb{I}$) for $\mathcal{H}_{\rm AM}$.
The Hamiltonian $\mathcal{H}^{(\text{II})} $ of model II is given by 
\begin{align}
\mathcal{H}^{(\text{II})} & = 
U \sum_{\alpha={\rm A},{\rm B}} n_{i_0\alpha\uparrow} n_{i_0\alpha\downarrow} 
+ U^{\prime} n_{i_0{\rm A}} n_{i_0{\rm B}} \nonumber \\
& - 2 J \left( {\bm S}_{i_0{\rm A}} \cdot {\bm S}_{i_0{\rm B}} + \frac{1}{4} n_{i_0{\rm A}} n_{i_0{\rm B}} \right) \nonumber \\
& + J^{\prime} 
\left( 
c_{i_0{\rm A}\uparrow}^{\dagger} c_{i_0{\rm A}\downarrow}^{\dagger} c_{i_0{\rm B}\downarrow} c_{i_0{\rm B}\uparrow} +
c_{i_0{\rm B}\uparrow}^{\dagger} c_{i_0{\rm B}\downarrow}^{\dagger} c_{i_0{\rm A}\downarrow} c_{i_0{\rm A}\uparrow}
\right) \nonumber \\
& 
- \frac{1}{2}(U+2U^{\prime}-J) \sum_{\alpha={\rm A},{\rm B}} n_{i_0 \alpha} \nonumber \\
& - t  \sum_{\langle i, j \rangle} \sum_{ s=\uparrow,\downarrow}
( c_{i \pi  s}^{\dagger} c_{j \pi  s} + c_{j \pi  s}^{\dagger} c_{i \pi  s} ) \nonumber \\
& - t_{\pi}  \sum_{\mbox{\scriptsize{$\llangle$}} i, j \mbox{\scriptsize{$\rrangle$}}} \sum_{ s=\uparrow,\downarrow}
( c_{i \pi  s}^{\dagger} c_{j \pi  s} + c_{j \pi  s}^{\dagger} c_{i \pi  s} ),  
\label{eq:refmodel:II}
\end{align}
where $c_{i_0\alpha s}$ ($c_{i_0\alpha s}^{\dagger}$) is the annihilation (creation) operator 
of an electron at the impurity site $\alpha\,(={\rm A},{\rm B})$ with spin $s\,(=\uparrow,\downarrow)$, 
$n_{i_0\alpha s}=c^\dag_{i_0\alpha s}c_{i_0\alpha s}$, and  
$n_{i_0\alpha} = n_{i_0\alpha \uparrow} + n_{i_0\alpha \downarrow}$ with  
$i_0$ labeling the location of the impurity site B in the honeycomb lattice as well as the impurity site A. 
$c_{i\pi s}$ ($c_{i\pi s}^{\dagger}$) is the annihilation (creation) operator of an electron at conduction site $i$ with 
spin $s\,(=\uparrow,\downarrow)$ and $c_{i_0\pi s} \equiv c_{i_0{\rm B} s}$. 
${\bm S}_{i_0\alpha}=(S_{i_0\alpha}^x, S_{i_0\alpha}^y, S_{i_0\alpha}^z)$ is 
the spin operator at the impurity site $\alpha$ given by 
\begin{equation}
S_{i_0\alpha}^{\nu} = \frac{1}{2} \sum_{s,s^{\prime}=\uparrow,\downarrow} 
c_{i_0\alpha s}^{\dagger} [ \mbox{\boldmath{$\sigma$}}^{\nu} ]_{s s^{\prime}} c_{i_0\alpha s^{\prime}}. \quad 
(\nu=x,y,z)
\label{eq:spinoperator_imp}
\end{equation}
The sums indicated by $\langle i,j\rangle$ and $\llangle i, j \rrangle$ run over all pairs 
of nearest and next nearest pairs of sites $i$ and $j$, respectively, in the honeycomb lattice, including the impurity site B. 
Notice that the fourth term in Eq.~(\ref{eq:refmodel:II}) 
is introduced to correct the double counting of the interactions, as in the case of $\mathcal{H}_{\rm AM}$.

Using the Lanczos transformation for the single-particle hopping terms in the honeycomb lattice, 
model I described by the Hamiltonian $\mathcal{H}^{(\text{I})}$ is mapped onto 
the following Q1D model: 
\begin{align}
\tilde{\mathcal{H}}^{(\text{I})}_{\rm Q1D} & = 
U \tilde{n}_{1 \uparrow} \tilde{n}_{1 \downarrow} 
- \frac{U}{2} (\tilde{n}_{1 \uparrow} + \tilde{n}_{1 \downarrow}) \nonumber \\
& 
+ V \sum_{ s=\uparrow,\downarrow} 
( \tilde{c}_{1  s}^{\dagger} \tilde{c}_{2  s} + \tilde{c}_{2  s}^{\dagger} \tilde{c}_{1  s} )
\nonumber \\
& + 
\sum_{l=2}^{L} \sum_{ s=\uparrow,\downarrow}
\varepsilon_l 
\tilde{c}_{l  s}^{\dagger} \tilde{c}_{l  s} 
\nonumber \\
& +
\sum_{l=2}^{L-1} 
\sum_{ s=\uparrow,\downarrow} 
t_l 
( \tilde{c}_{l  s}^{\dagger} \tilde{c}_{l+1  s} 
+ \tilde{c}_{l+1  s}^{\dagger} \tilde{c}_{l  s} ), 
\label{eq:refmodel:q1d:I} 
\end{align}
where $\tilde{c}_{l  s}$ ($\tilde{c}_{l s}^{\dagger}$) is  
the electron creation (annihilation) operator generated by the $(l-1)$th Lanczos iteration $(l\ge2)$
with $\tilde{c}_{1  s} \equiv c_{{\rm H}  s}$ and  
$\tilde{c}_{2  s} \equiv c_{i_0 \pi  s}$, $\tilde{n}_{l  s} = \tilde{c}_{l s}^{\dagger} \tilde{c}_{l s}$, and 
the generalized Lanczos coefficients 
$\varepsilon_l $ and $t_l$ are determined through the Lanczos procedure~\cite{shirakawa1}. 
The schematic depiction of model I in the Q1D representation described by the Hamiltonian $\tilde{\mathcal{H}}^{(\text{I})}_{\rm Q1D}$ 
is shown in Fig.~\ref{fig:repmodel}(c).

Similarly, model II described by the Hamiltonian $\mathcal{H}^{(\text{II})}$ is mapped onto 
the following Q1D model: 
\begin{align}
\tilde{\mathcal{H}}^{(\text{II})}_{\rm Q1D} & = 
U \sum_{l=1,2} \tilde{n}_{l\uparrow} \tilde{n}_{l\downarrow} 
+ U^{\prime} \tilde{n}_{1} \tilde{n}_{2} \nonumber \\
& - 2 J \left( \tilde{\bm S}_{1} \cdot \tilde{\bm S}_{2} + \frac{1}{4} \tilde{n}_{1} \tilde{n}_{2} \right) \nonumber \\
& + J^{\prime} 
\left( 
\tilde{c}_{1\uparrow}^{\dagger} \tilde{c}_{1\downarrow}^{\dagger} \tilde{c}_{2\downarrow} \tilde{c}_{2\uparrow} +
\tilde{c}_{2\uparrow}^{\dagger} \tilde{c}_{2\downarrow}^{\dagger} \tilde{c}_{1\downarrow} \tilde{c}_{1\uparrow}
\right) \nonumber \\
& 
- \frac{1}{2}(U+2U^{\prime}-J) \sum_{l=1,2} \tilde{n}_{l} \nonumber \\
& 
+ 
\sum_{l=2}^{L}
\sum_{ s=\uparrow,\downarrow} 
\varepsilon_l 
\tilde{c}_{l  s}^{\dagger} \tilde{c}_{l  s} 
\nonumber \\
& +
\sum_{l=2}^{L-1} 
\sum_{ s=\uparrow,\downarrow} 
t_l 
( \tilde{c}_{l  s}^{\dagger} \tilde{c}_{l+1  s} 
+ 
\tilde{c}_{l+1  s}^{\dagger} \tilde{c}_{l  s} ),
\label{eq:refmodel:q1d:II}
\end{align}
where $\tilde{n}_l = \tilde{n}_{l\uparrow} + \tilde{n}_{l\downarrow}$, 
$\tilde{\bm S}_l$ is the spin operator given in Eq.~(\ref{eq:spinoperator_imp}) 
but for $\tilde{c}_{l s}$ instead of $c_{i_0\alpha  s}$, 
$\tilde{c}_{1 s} \equiv c_{i_0 {\rm A} s}$, and 
$\tilde{c}_{2 s} \equiv c_{i_0 {\rm B} s}$. 
The schematic depiction of model II in the Q1D representation described by the Hamiltonian $\tilde{\mathcal{H}}^{(\text{II})}_{\rm Q1D}$ 
is shown in Fig.~\ref{fig:repmodel}(d). 
Notice that the generated Lanczos coefficients $\varepsilon_l$ and $t_l$ 
in Eq.~(\ref{eq:refmodel:q1d:II}) are exactly the same as those in Eq.~(\ref{eq:refmodel:q1d:I}). 
Therefore, the difference between models I and II given in Eqs.~(\ref{eq:refmodel:q1d:I}) and (\ref{eq:refmodel:q1d:II}) 
is a type of couplings between sites $l=1$ and 2 represented by operators $\tilde{c}_{1 s}$ and $\tilde{c}_{2 s}$, respectively: 
these two sites are connected through the hopping in model I while 
they are connected through the interactions in model II,
as also indicated in Figs.~\ref{fig:repmodel}(c) and \ref{fig:repmodel}(d), 
besides the site at $l=2$ being one of the impurity sites in model II.

Model I is called a pseudogap Anderson model, 
in which the impurity site is connected to the conduction band 
through a hybridization function with a pseudogap structure at Fermi level, 
corresponding to the undoped case in model I when the chemical potential $\mu$ is at $ 2 t_{\pi}\,(\equiv \mu_0)$. 
The ground state of model I in the undoped case 
is in the local moment phase where a free magnetic moment exists~\cite{shirakawa1}.
In the doped case with $\mu \neq \mu_0$, the ground state 
is in either the Kondo screening phase (i.e., the symmetric strong coupling phase) 
or the asymmetric strong coupling phase,  
where no free magnetic moment appears~\cite{vojta}.

\begin{figure}[htbp]
\begin{center}
\includegraphics[width=\hsize]{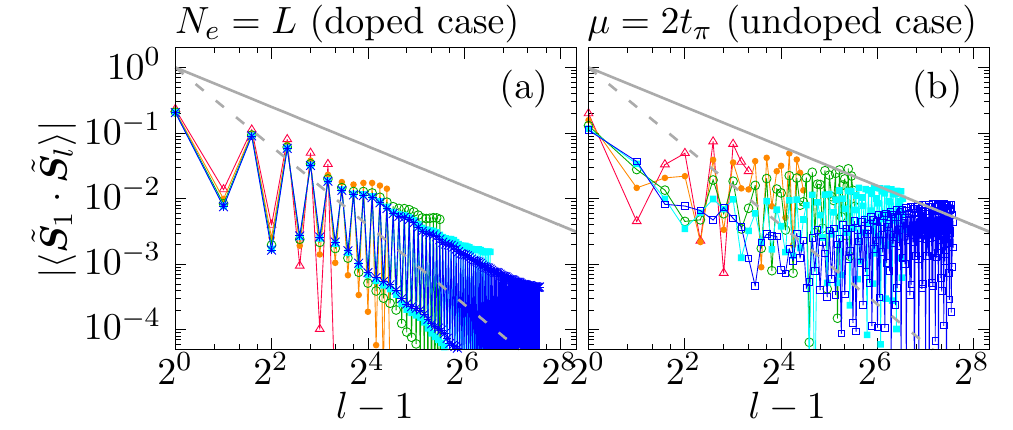}
\caption{
Log-log plots of the absolute values of the 
spin correlation function $\langle \tilde{{\bm S}}_{1} \cdot \tilde{\bm S}_{l} \rangle$ between the impurity and conduction sites 
for (a) the doped case and (b) the undoped case with various values of $L$ in the Q1D model $\tilde{\mathcal{H}}_{\rm Q1D}^{\rm (I)}$.  
The number $N_e$ of electrons is set to be $N_e=L$ in (a), corresponding to the doped case, 
while $N_e$ is adjusted so as to minimize $E_0(N_e)- \mu_0 N_e$ with the chemical potential $\mu_0 = 0.25 t_{\pi}$ in (b), 
corresponding to the undoped case, where $E_0(N_e)$ is the ground state energy of $\tilde{\mathcal{H}}_{\rm Q1D}^{\rm (I)}$ 
with $N_e$ electrons in the one-dimensional cluster of size $L$. 
The other parameters used are $U=3t$, $V=t$, and $t_{\pi} = t/12$. 
Red, orange, green, cyan, and blue symbols 
correspond to the results for $L=12$, $24$, $48$, $96$, and $192$, respectively. 
Solid and dashed lines represent functions proportional to $(l-1)^{-1}$ and $(l-1)^{-2}$, respectively, for comparison. 
}
\label{fig:repspin}
\end{center}
\end{figure}

Figure~\ref{fig:repspin} shows the spin correlation function 
$\langle \tilde{\bm S}_1 \cdot \tilde{\bm S}_l \rangle$ 
between the impurity site ($l=1$) and the conduction sites ($l\ge2$) in the Q1D representation of the 
single-impurity Anderson model $\tilde{\mathcal{H}}^{(\text{I})}_{\rm Q1D} $. 
We find in Fig.~\ref{fig:repspin}(a) that the spin correlation function $\langle \tilde{\bm S}_1 \cdot \tilde{\bm S}_l \rangle$
for the doped case decays faster than $d^{-1}$, where $d = l-1$ is the ditance between the impurity site and the 
conduction site labeled by $l$ ($l\ge2$) in the Q1D model $\tilde{\mathcal{H}}^{(\text{I})}_{\rm Q1D} $.
In contrast, $\langle \tilde{\bm S}_1 \cdot \tilde{\bm S}_l \rangle$ for 
the undoped case in Fig.~\ref{fig:repspin}(b) shows somewhat nonmonotonic behavior 
as a function of the distance $d$. If we focus on the distances near the impurity site  
in the largest system (denoted by blue squares), 
$\langle \tilde{\bm S}_1 \cdot \tilde{\bm S}_l \rangle$ decays approximately as $d^{-1}$.
If we focus on the spin correlation function at the maximum distance of a given system size $L$, 
i.e., $\langle \tilde{\bm S}_1 \cdot \tilde{\bm S}_L \rangle$, 
the envelope seems to decay slower that $d^{-1}$. 
Although, it is not easy to determine the exponent precisely in Fig.~\ref{fig:repspin}(b),  
we can safely claim the much slower decay 
of $\langle \tilde{\bm S}_1 \cdot \tilde{\bm S}_l \rangle$ 
for the undoped case than for the doped case.

\begin{figure}[htbp]
\begin{center}
\includegraphics[width=\hsize]{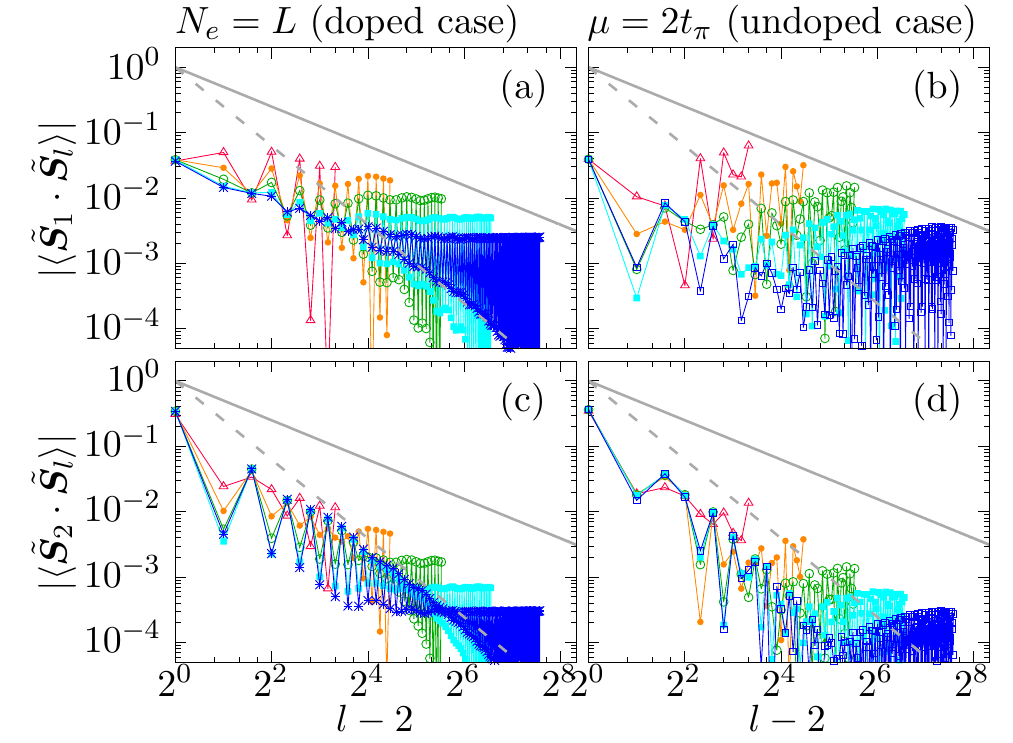}
\caption{
Log-log plots of the absolute values of the 
spin correlation functions (a,b) $\langle \tilde{{\bm S}}_{1} \cdot \tilde{\bm S}_{l} \rangle$ 
and (c,d) $\langle \tilde{{\bm S}}_{2} \cdot \tilde{\bm S}_{l} \rangle$ between the impurity and conduction sites 
for the doped case (left panels) and the undoped case (right panels) with various values of $L$ in the Q1D model 
$\tilde{\mathcal{H}}_{\rm Q1D}^{\rm (II)}$. 
The number $N_e$ of electrons is set to be $N_e=L$ in (a,c), corresponding to the doped case, 
while $N_e$ is adjusted so as to minimize $E_0(N_e)- \mu_0 N_e$ with the chemical potential $\mu_0 = 0.25 t_{\pi}$ in (b,d), 
corresponding to the undoped case, where $E_0(N_e)$ is the ground state energy of $\tilde{\mathcal{H}}_{\rm Q1D}^{\rm (II)}$ 
with $N_e$ electrons in the one-dimensional cluster of size $L$. 
The other parameters used are $U=3t$, $J = 0.1 U$, and $t_{\pi} = t/12$. 
Red, orange, green, cyan, and blue symbols 
correspond to the results for $L=12$, $24$, $48$, $96$, and $192$, respectively. 
Solid and dashed lines represent functions proportional to $(l-2)^{-1}$ and $(l-2)^{-2}$, respectively, for comparison. 
}
\label{fig:repspin2}
\end{center}
\end{figure}

In model II, the local spin triplet is favored at the impurity sites 
at $l=1$ and $2$ because of the Hund's coupling in the strong coupling phase (e.g., with 
$U=3t$, $J = 0.1 U$, and $t_{\pi} = t/12$) for the undoped case, 
and this spin-1 impurity is coupled to the conduction band
via the hybridization function that diverges at Fermi level~\cite{shirakawa1}. 
The Fermi level is shifted away from this diverging point of the hybridization 
function in the doped system. However, in both undoped and doped cases, 
the impurity spin is similarly screened partially by the conduction electrons and 
there appears residual unscreened free magnetic moment with its spin as large as 1/2~\cite{mattis1967,nozieres1980}.

Figures~\ref{fig:repspin2}(a) and \ref{fig:repspin2}(b) show the spin correlation function 
$\langle \tilde{\bm S}_1 \cdot \tilde{\bm S}_l \rangle$ 
between the impurity site ($l=1$) and the conduction sites ($l\ge3)$. 
We find that in both undoped and doped cases 
$\langle \tilde{\bm S}_1 \cdot \tilde{\bm S}_l \rangle$ decays rather close to $d^{-1}$ with the distance $d=l-2$
and shows a similar behavior to that for the undoped case in model I [see Fig.~\ref{fig:repspin}(b)],  
where the free magnetic moment appears. 
In contrast, as shown in Figs.~\ref{fig:repspin2}(c) and \ref{fig:repspin2}(d), 
the spin correlation function 
$\langle \tilde{\bm S}_2 \cdot \tilde{\bm S}_l \rangle$ 
from the second impurity site ($l=2$), which is directly 
connected to the conduction sites, decays approximately as $d^{-2}$,
specially focusing on the region close to the impurity site
for the larger systems. This $d^{-2}$ decay is expected for the Fermi liquid 
and thus it is consistent with the partial screening of the impurity spin at $l=2$.

When we compare these results with those for the Anderson model $\mathcal{H}_{\rm AM}$ in Sec.~\ref{sec:spin}, 
we have to remind that the accessible system size is very limited for the Anderson model $\mathcal{H}_{\rm AM}$. 
For example, the nearly $d^{-2}$ decay of the spin correlation functions 
$\langle \tilde{\bm S}_1 \cdot \tilde{\bm S}_l \rangle$ in Fig.~\ref{fig:repspin}(a) and 
$\langle \tilde{\bm S}_2 \cdot \tilde{\bm S}_l \rangle$ in Figs.~\ref{fig:repspin2}(c) 
and \ref{fig:repspin2}(d) can still be observed even in the moderate systems as large as $L=48$, 
which is the largest accessible system size for the Anderson model $\mathcal{H}_{\rm AM}$ in Sec.~\ref{sec:spin}. 
Indeed, we can observe the spin correlation function consistent with the $d^{-2}$ behavior in 
Figs.~\ref{fig:spinspin2}(a), \ref{fig:spinspin2}(c), and \ref{fig:spinspin2}(d). 
On the other hand, the spin correlation function $\langle \tilde{\bm S}_1 \cdot \tilde{\bm S}_l \rangle$ 
for $\mathcal{H}^{({\text {II}})}_{\rm Q1D}$ 
with $L=48$, denoted by green symbols in Figs.~\ref{fig:repspin2}(a) and \ref{fig:repspin2}(b), 
appears to be constant rather than decaying algebraically with some power, which is similar to the behavior of  
the spin correlation function $\langle {\bm S}_{\sigma} \cdot \tilde{\bm S}_{lc} \rangle$ ($c = a,e$) 
in Fig.~\ref{fig:spinspin}. 

The difference between the doped and undoped cases in these systems is most difficult to distinguish, 
since the exponent for the undoped case deviates from not only $d^{-1}$ 
but also $d^{-2}$ even in the largest system of model I, as shown in Fig.~\ref{fig:repspin}(b). 
Rather, the exponent seems to vary between $d^{-1}$ and $d^{-2}$ for different system sizes.  
Therefore, the clear determination of the decay exponent 
is difficult even in the simplest model such as model I, based on the results available at this moment. 
However, we can assert that the nontrivial decay exponent slower than $d^{-2}$ 
is the hallmark of the presence of the free magnetic moment~\cite{mitchell2}.

\bibliography{vac_graphene}

\end{document}